\documentclass[11pt,letterpaper]{amsart}
\usepackage[T1]{fontenc}
\usepackage{lmodern}
\usepackage{xcolor}
\usepackage{amssymb,mathtools,mathrsfs}
\usepackage{bm}
\usepackage{microtype,booktabs,enumitem}
\usepackage[hidelinks]{hyperref}
\hypersetup{pdftitle={The concentration conjecture for Dirac-Coulomb operators},
 pdfauthor={Habib Ammari and Jiayu Qiu},pdfsubject={Sharp spectral comparison for Coulomb charge distributions},
 pdfkeywords={Dirac-Coulomb, concentration conjecture, spectral gap, weighted resolvent}}
\numberwithin{equation}{section}
\newtheorem{theorem}{Theorem}[section]
\newtheorem{lemma}[theorem]{Lemma}
\newtheorem{proposition}[theorem]{Proposition}
\newtheorem{corollary}[theorem]{Corollary}
\theoremstyle{remark}
\newtheorem{remark}[theorem]{Remark}
\newcommand{\R}{\mathbb R}
\newcommand{\C}{\mathbb C}
\DeclareMathOperator{\Dom}{Dom}

\title[Concentration for Dirac-Coulomb operators]
{The concentration conjecture for Dirac-Coulomb operators}
\author{Habib Ammari}
\address{Department of Mathematics, ETH Z\"urich,
R\"amistrasse 101, 8092 Z\"urich, Switzerland}
\email{habib.ammari@math.ethz.ch}
\author{Jiayu Qiu}
\address{Department of Mathematics, ETH Z\"urich,
R\"amistrasse 101, 8092 Z\"urich, Switzerland}
\email{jiayu.qiu@math.ethz.ch}
\thanks{Corresponding author: Jiayu Qiu.}

\date{}

\begin{document}

\begin{abstract}
We prove the concentration conjecture of Esteban-Lewin-S\'er\'e for three-dimensional Dirac-Coulomb operators. Let $D_m$ be the free Dirac operator of mass $m>0$, and let $\mu$ be a nonnegative finite measure on $\R^3$ with total charge $G\in(0,1]$. The main result is that the lowest eigenvalue of $D_m-\mu*|\cdot|^{-1}$ in the gap $(-m,m)$ is at least $m\sqrt{1-G^2}$, with equality precisely when the entire charge is concentrated at one point. The proof is based on the Birman-Schwinger formulation, which reduces the eigenvalue estimate and its equality case to a strict two-center estimate for the free Dirac resolvent. In particular, we show that for all energies $0\leq E<m$, the norm of resolventness with Coulomb weights at different centers is strictly less than its value of one-center $m/\sqrt{m^2-E^2}$. This estimate is proved by an axial reduction in prolate spheroidal coordinates, combining a scalar resolvent comparison at large separation with an angular Schur complement at small separation.
\end{abstract}

\maketitle

\medskip

\tableofcontents

\section{Introduction and main result}

Let $m>0$ and consider the free Dirac operator on $\R^3$,
\begin{equation}\label{intro:dirac}
 D_m=-i\boldsymbol\alpha\cdot\nabla+m\beta
 \quad\text{in }\mathcal H=L^2(\R^3;\C^4).
\end{equation}
The Hermitian matrices $\alpha_1,\alpha_2,\alpha_3,\beta$ satisfy the Clifford relations. Given distinct points $A_1,\ldots,A_N\in\R^3$, with $N\geq 2$, and charges $g_1,\ldots,g_N>0$, let
\begin{equation}\label{intro:potential}
 V(x)=\sum_{i=1}^N\frac{g_i}{|x-A_i|},\quad
 G=\sum_{i=1}^N g_i \in (0,1].
\end{equation}
Since $N\geq2$ and $G\leq1$, every individual charge satisfies $g_i<1$. We denote by $H_V$ the unique self-adjoint realization of $D_m-V$ whose domain is contained in $H^{1/2}(\mathbb R^3;\mathbb C^4)$, as in \cite{ELSI}. For the one-center case ($N=1$), $H_V$ is defined as the distinguished Coulomb realization. In particular, at the critical charge $G=1$, the distinguished operator is the norm-resolvent limit of the subcritical realizations as the charge increases to one. Its domain is then no longer contained in $H^{1/2}$; see \cite[Theorem~1.11]{ELScritical2019}. For finite configurations it therefore suffices to consider $N\geq 2$.

It is known that $H_{V}$ has the same essential spectrum $(-\infty,-m]\cup[m,\infty)$ as the free Dirac operator $D_{m}$, and its spectrum in the gap \((-m,m)\) is discrete; see \cite[Theorem 2.1]{ELSI}. Moreover, for the
configurations in \eqref{intro:potential},  writing \(\lambda_1(H_V)\) for the
lowest eigenvalue in the gap, Theorem \ref{thm:main} below shows that it is strictly positive.  Physically, $\lambda_1(H_{V})$ refers to the lowest energy of a bound relativistic electron residing in the $N$-center attractive Coulomb potential, with the interaction parameter being $G$.

With these notions, the concentration conjecture states that for fixed $G$, the smallest $\lambda_1(H_{V})$ occurs when the centers $A_i$ coincide, i.e., $N=1$. For the latter case, it is known that $\lambda_1(H_{V})=m\sqrt{1-G^2}$ (see \eqref{pre:atomic-energy} and the references there). The main result of this paper gives a definitive answer to this conjecture.

\begin{theorem} \label{thm:main}
Let $m>0$, $N\geq 2$, $g_i>0$, and $A_i\in\R^3$ be pairwise distinct. Suppose
$0<G\leq1$. Then $H_V$
has a lowest eigenvalue in $(-m,m)$, and
\begin{equation}\label{intro:main}
 \lambda_1(H_V)> m\sqrt{1-G^2} .
\end{equation}
\end{theorem}
The two-center estimate also gives a direct proof for arbitrary nonnegative charge measures as pointed out in \cite{ELSsurvey}. For completeness, we give the proof in Section~\ref{sec:conclusion}.
\begin{corollary}  \label{cor:general-charges}
Let $m>0$ and let $\mu$ be a nonnegative finite Borel measure on $\R^3$ with $G:=\mu(\R^3)\in(0,1]$. Define
\[
V_\mu(x):=\int_{\R^3}\frac{d\mu(y)}{|x-y|}.
\]
Let $H_\mu$ be the distinguished self-adjoint realization of $D_m-V_\mu$. Then $H_\mu$ has a lowest eigenvalue $\lambda_1(H_\mu)$ in $(-m,m)$, which satisfies
\begin{equation}\label{intro:general-charges}
 \lambda_1(H_\mu)\geq  m\sqrt{1-G^2}.
\end{equation}
The equality holds if and only if \(\mu = G \delta_A\) for some \(A \in \mathbb{R}^3\). At $G = 1$, the single-center operator is understood in the critical distinguished sense.
\end{corollary}

The multi-center concentration problem dates back to Klaus's work in 1980 \cite{Klaus} for the two-center case and was subsequently investigated by Briet and Hogreve \cite{BH}, who proved the united-atom lower bound for two equal, sufficiently small charges. Bosi, Dolbeault and Esteban \cite{BDE2008} obtained related estimates for Dirac operators with several Coulomb singularities using multipolar Hardy inequalities. Numerical calculations for the two-center problem, including the Cassini-coordinate computations of Artemyev et al. \cite{Artemyev2010}, provide further evidence for this concentration; see also \cite{ELSsurvey}. Esteban, Lewin and S\'er\'e formulated the conjecture for general charge distributions in \cite{ELSI,ELSII} and surveyed it in \cite{ELSsurvey}. Their first paper \cite{ELSI} constructs the distinguished realization and gives min--max formulas for its gap eigenvalues. In \cite{ELSII}, the authors study the optimization of the first gap eigenvalue over charge distributions of fixed mass and relate the problem to universal critical coupling constants and Hardy-Dirac inequalities. Below the critical charge, they prove the existence of an optimal measure and show that every optimizer is supported on a compact set of zero Lebesgue measure. These works established a framework for the concentration problem, and it remains to identify the point charge as the unique optimizer.

The interest in the multi-center concentration conjecture lies not only in its mathematical importance but also in the physical implications in nuclear geometry and the relativistic electronic binding problem. It predicts that among all configurations with a prescribed total charge, the united atom should produce the lowest electronic energy. In the non-relativistic one-electron theory, where the concerned operator becomes Schrödinger-type, the multi-center concentration is well justified, which follows from the variational principle and concavity of the ground-state energy in the charge distribution. Lieb and Simon \cite{LiebSimon1978} proved monotonicity in the internuclear separation for two nuclei. Lieb \cite{Lieb1982} established the contraction comparison for general finite configurations. In contrast, the relativistic problem is substantially different: the Dirac operator is unbounded from below, and its eigenvalues in the spectral gap are characterized by min-max principles; see \cite{DES2000} and, for general charge distributions, \cite{ELSI,ELSsurvey}. Consequently, the elementary concavity argument does not extend directly to this setting. To overcome this difficulty, we apply a Birman-Schwinger formulation, with the main idea summarized in the following subsection.

\subsection{Main idea of proof}

Let us focus on the finite configurations in \eqref{intro:potential}. The idea of the proof to Theorem \ref{thm:main} is to show that the associated Birman-Schwinger equation has no solution within $E\in [0,m\sqrt{1-G^2}]$, and therefore, the lowest positive eigenvalue lies strictly above the claimed lower bound. To be precise, we will show that $1\notin \text{spec}(\mathcal{K}_{E})$ for $E\in [0,m\sqrt{1-G^2}]$, where $\mathcal{K}_{E}$ is the Birman-Schwinger operator:
\begin{equation*}
\mathcal K_E:=WR_EW^*
\end{equation*}
with $W:\psi\mapsto (\sqrt{g_i}\,w_{A_i}\psi)_{i=1}^N$ (vectorization of the potential), $w_A(x)=|x-A|^{-1/2}$ (the potential weight), and $R_E=(D_m-E)^{-1}$ (the free resolvent). Hence, proving $1\notin \text{spec}(\mathcal{K}_{E})$ reduces to the spectral estimate of an $N\times N$ matrix, whose diagonal element corresponds to the single-center-potential Dirac-Coulomb operator, and it is known that
\begin{equation}\label{intro:atomic-norm}
 \|w_A R_Ew_A\|=\frac m\kappa ,\quad \text{with }\,\kappa:=\sqrt{m^2-E^2}.
\end{equation}
See \cite{CPV}. Importantly, our central estimate is on the off-diagonal element:
\begin{theorem} \label{thm:cross}
For $m>0$, $0\leq E<m$, and distinct $A,B\in\R^3$,
\begin{equation}\label{intro:cross}
 \bigl\|w_B(D_m-E)^{-1}w_A\bigr\|
 <\frac{m}{\sqrt{m^2-E^2}}.
\end{equation}
\end{theorem}
We refer the reader to Section \ref{sec:conclusion} for details on how Theorem \ref{thm:cross} leads to the conclusion of Theorem \ref{thm:main}. In fact, Theorem \ref{thm:main} is inspired by the intuition that when the centers are well-separated, i.e., $|A-B|\to\infty$, the operator norm in \eqref{intro:cross} decays to zero by the decay of potentials as well as that of the free resolvent when $|E|<m$. On the other hand, for any small separation, the operator norm should be strictly smaller than the exact identity \eqref{intro:atomic-norm}.

The major content of this paper is dedicated to justifying the above intuition, especially on two estimates whose ranges together cover all positive separations.
To this end, we first relate the singular value of $T_{E}:=w_B(D_m-E)^{-1}w_A$ to the eigenvalue of an explicit auxiliary operator $H^{(3)}_{k,q,j}$ with $k=\kappa|A-B|/2$ identifying the separation of centers, $q$ linked to the original energy, and $j$ representing the corresponding sector of axial angular momentum (along the line joining $A,B$); see Section \ref{sec:axial} for details. This auxiliary operator provides a unified framework for dealing with both the large- and small-separation regimes. Specifically, in Proposition \ref{ax:singular-vector}, we prove that a violation of \eqref{intro:cross} would produce an eigenvalue $0<\lambda\leq1$ of $H^{(3)}_{k,q,j}$ at $q=(E/m)\lambda\in[0,1]$. 

With these observations, it therefore suffices to prove that $[-1,1]$ forms a spectral gap of the auxiliary operator $H^{(3)}_{k,q,j}$ for all $k>0$ and $q\in[0,1]$. This is first proved for the range $k\geq2/5$ in Section \ref{end:section}, i.e., the large-separation regime. More precisely, we start by a zero-energy estimate (i.e., $q=0$), which is facilitated by the observation that the square $(H^{(3)}_{k,0,j})^2$ decomposes perfectly into the sum of a radial and an angular operator. Then, a scalar resolvent estimate, together with a concavity argument, establishes the desired spectral gap for all $q\in[0,1]$ when $k\geq2/5$.

Next, for smaller $k$, we proceed by a perturbation argument. In particular, we show in Section \ref{sec:radial} that the auxiliary operator $H^{(3)}_{k,q,1/2}$ (the lowest angular sector suffices to prove the gap estimate) admits a near-perfect decomposition as mentioned above, except for a small perturbation proportional to $k$; see \eqref{rad:decomposition}. In particular, the radial operator appearing in the decomposition has a two-dimensional kernel and an explicit positive spectral gap. This provides a Schur complement formalism adapted to the lowest positive angular mode in this kernel, as detailed in Section~\ref{sec:angular}. Since the Schur complement is scalar, it gives access to obtain the gap estimate by explicit (yet complicated) calculations; see Section~\ref{sec:bridge}. We isolate all the exact scalar verifications in Appendix~\ref{app:scalar} to guarantee a smooth understanding of the main idea of proof.

\section{Notation and preliminaries}\label{sec:preliminaries}

\subsection{Notation and conventions}
\label{subsec:notation}

In this paper, all Hilbert spaces are complex, and their inner products are linear in the second argument. We omit the domain and the value space in function-space notation if no confusion arises. For example, sometimes we write
\[
 L^2=L^2(\mathbb R^3;\mathbb C^4)=\mathcal H,
 \quad H^s=H^s(\mathbb R^3;\mathbb C^4) \quad (s\in\mathbb R).
\]
Here, $H^s$ denotes the usual $L^2$-based Sobolev space; negative-order spaces on $\mathbb R^3$ are understood by duality through the $L^2$ pairing. The same abbreviations will be applied for different purposes, e.g. $L^2$ denoting $L^2(\mathbb R^3;\mathbb C)$ when considering a scalar resolvent, which will be clear from the context. Unless another measure is explicitly displayed, these function spaces are defined with respect to the Lebesgue measure in the indicated variables.

We generally omit subscripts from norms and inner products
when the underlying spaces are clear. Taking the functions in
$L^2(U;\mathbb C^d)$ for example ($U$ is an open domain), we write
\[
 \langle f,g\rangle=\int_U f(x)^*g(x)\,dx, \quad
 \|f\|^2=\int_U |f(x)|^2\,dx,
\]
where $|f(x)|$ is the Euclidean length of the pointwise vector. For $z\in\mathbb C^d$, $\|z\|$ denotes the Euclidean norm. Finite direct sums of Hilbert spaces carry the norm
\[
 \|(f_1,\ldots,f_N)\|^2=\sum_{i=1}^N\|f_i\|^2.
\]
In particular, $\mathcal H^N$ denotes the direct sum of $N$ copies of $\mathcal H=L^2(\mathbb R^3;\mathbb C^4)$. For a bounded operator $T:\mathscr X\to\mathscr Y$ between Hilbert spaces $\mathscr X$ and $\mathscr Y$, the same notation denotes the induced operator norm,
\[
 \|T\|:=\sup_{\substack{f\in\mathscr X\\\|f\|=1}}\|Tf\|.
\]
For a closed operator $T$, its graph norm is
\[
 \|f\|_T^2:=\|f\|^2+\|Tf\|^2,
 \qquad f\in\Dom(T).
\]
We use $\Dom(T)$, $\ker T$, $\operatorname{Ran}T$, and $\operatorname{spec}(T)$ for the domain, kernel, range, and spectrum of an operator. An asterisk $*$ denotes the Hilbert-space adjoint, or the corresponding duality adjoint when Sobolev spaces are involved; for matrices, it denotes conjugate transpose. The rank-one operator associated with a vector is denoted by the bra-ket notation:
\[
 (|u\rangle\langle v|)f:=u\langle v,f\rangle.
\]

The identity operator is denoted by $I$, with its space being clear from context. Tensor-product symbols and identity factors are omitted when no confusion arises; operators acting on distinct tensor factors commute. We write the commutator and anti-commutator as
\[
 [A,B]:=AB-BA,\qquad \{A,B\}:=AB+BA.
\]

\subsection{The free Dirac operator, Coulomb potential, and the Dirac-Coulomb operator}
Note that the matrices in the free Dirac operator $D_m$ satisfy the Clifford relations:
\[
 \alpha_i\alpha_j+\alpha_j\alpha_i=2\delta_{ij}I,
 \qquad\alpha_i\beta+\beta\alpha_i=0,\qquad\beta^2=I.
\]
It is well-known that 1) $D_m$ is self-adjoint on $H^1(\R^3;\C^4)$, 2) $D_m^2=(-\Delta+m^2)I$, and 3) $\text{spec}(D_{m})=(-\infty,-m]\cup[m,\infty)$; see \cite[Section 1.4]{Thaller1992}. Consequently, the free resolvent is rewritten as
\begin{equation}\label{pre:free-factorization}
 R_E=(D_m+E)(-\Delta+\kappa^2)^{-1},\quad |E|<m.
\end{equation}

The square-root potential weight $w_A=|x-A|^{-1/2}$ will be frequently applied throughout this paper, with $A\in\mathbb{R}^3$. Note that, as a multiplication operator, $w_{A}$ maps $H^{1/2}$ continuously into $L^2$ by the fractional Hardy inequality (cf. \cite[Eq. (4.3)]{ELSI}). Thus, its adjoint maps $L^2$ into $H^{-1/2}$; with a little abuse of notation, we will denote the adjoint of $w_{A}$ by the same symbol. Since the free resolvent $R_E$ maps $H^{-1/2}$ into $H^{1/2}$, these facts define the bounded weighted extensions
\begin{equation*}
w_B R_Ew_A:\, L^2\to L^2.
\end{equation*}
When $A=B$, the sharp single-center estimate \eqref{intro:atomic-norm} follows from the weighted Hardy--Dirac inequality of \cite[Theorem~1.1]{CPV}. In particular, for $0<E<m$, the equality is attained by the ground state of the atom with charge $g=\kappa/m$ that satisfies
\begin{equation*}
(D_m-g/|x-A|)\psi=E\psi,
\end{equation*}
by which it is clear that $w_A R_Ew_A\phi=g^{-1}\phi$ with $\phi:=w_A\psi$. At $E=0$, the norm in \eqref{intro:atomic-norm} follows by continuity as $E\downarrow0$, since the resolvent identity implies norm continuity of $R_E:H^{-1/2}\to H^{1/2}$. We also recall the explicit formula
\begin{equation}\label{pre:atomic-energy}
 \lambda_1(D_m-g/|x-A|)=m\sqrt{1-g^2},\qquad 0<g\leq1.
\end{equation}
For $0<g<1$, see \cite[Eq.~(1.17)]{CPV} and \cite[Section~7.4.2]{Thaller1992}. The critical case $g=1$ follows by norm-resolvent convergence as $g\uparrow1$ and spectral convergence on compact subintervals of the gap, as in \cite[Section~3.3]{ELSI}.

The construction of distinguished self-adjoint realizations of Dirac operators is closely related to Hardy–Dirac inequalities; see \cite{EL2007}. For general Coulomb charge distributions, we rely on the framework developed in \cite{ELSI}. For the case $N\geq 2$ so that each individual charge is smaller than one, \cite[Theorems~2.1 and~3.2]{ELSI} gives the distinguished self-adjoint realization $H_V$ with domain contained in $H^{1/2}$. Its domain is
\begin{equation}\label{pre:domain}
\Dom(H_V)=\bigl\{\psi\in H^{1/2}(\R^3;\C^4):
 (D_m-V)\psi\in L^2\text{ in distributions}\bigr\}.
\end{equation}
Indeed, $D_m-V:H^{1/2}\to H^{-1/2}$ is bounded and symmetric in the dual pairing. A spinor on the right of \eqref{pre:domain} therefore belongs to $\Dom(H_V^*)=\Dom(H_V)$ by testing against $\Dom(H_V)$. The reverse inclusion follows from the distributional action of $H_V$. The fractional Hardy inequality implies $V^{1/2}\psi\in L^2$ for every $\psi\in\Dom(H_V)$.

\section{The auxiliary axial operator}
\label{sec:axial}
As indicated in the Introduction, the estimate of the two-center weighted resolvent in Theorem~\ref{thm:cross} is reduced to proving the spectral gap of an auxiliary axial operator on the half-strip $\Omega:=(0,\infty)\times(0,\pi)$ with coordinates $(\mu,\nu)$. In this section, we explicitly construct this auxiliary operator and show how the singular values of the weighted resolvent $w_B(D_m-E)^{-1}w_A$ in Theorem~\ref{thm:cross} transfer to the spectrum of the auxiliary operator.

Throughout this section, we may assume without loss of generality that the two distinct centers are $A=(0,0,-a)$ and $B=(0,0,a)$, with $a>0$, by translation invariance and rotation covariance of the free Dirac operator. We also write
\[
 r_A=|x-A|,\quad r_B=|x-B|,\quad w_A=r_A^{-1/2},\quad w_B=r_B^{-1/2}.
\]
We will use three families of Pauli matrices, denoted by $\sigma_i$, $\zeta_i$ and $\tau_i$, with matrices from different families commuting. The matrices $\sigma_i$ and $\zeta_i$ act on two copies of $\mathbb C^2$, whose tensor product gives the physical spinor space $\mathbb C^4$ on which the Dirac operator acts. The matrices $\tau_i$ act on an additional $\mathbb C^2$ factor indexing a pair of physical spinors, so the doubled spinor space is $\mathbb C^2\otimes\mathbb C^4\simeq\mathbb C^8$. This doubling is natural when studying the singular values of $w_B(D_m-E)^{-1}w_A$, since a singular pair consists of two four-component spinors; see \eqref{eq:singular_doublet}. For ease of notation, we omit tensor-product symbols and identity factors whenever no confusion arises.

The axial angular momentum is the total angular momentum about the line joining the centers, i.e., the $z$-axis under the aforementioned convention. In cylindrical coordinates $(z,\rho,\varphi)$, its generator is
\[
 J_z:=L_z+S_z=-i\partial_\varphi+\frac12\Sigma_z,
 \quad \Sigma_z:=-i\alpha_x\alpha_y,
\]
where $L_z=-i(x\partial_y-y\partial_x)$ is the orbital part of the angular momentum and $S_z=\Sigma_z/2$ is the spin part. It is standard to check $\text{spec}(J_z)=\mathbb Z+\frac{1}{2}$; we write $\mathcal H_j=\ker(J_z-j)$ for the axial sector with the label $j\in\mathbb Z+\tfrac12$.

With this notation, we are ready to introduce the auxiliary axial operator and present its properties. For $k>0$, $q\in\mathbb R$ and $j\in\mathbb Z+\tfrac12$, we define the following operator on $C_c^\infty(\Omega;\mathbb C^8)$:
\begin{equation}\label{ax:pencil}
\begin{split}
H^{(3)}_{k,q,j}:=&-i\zeta_3\sigma_1\partial_\mu
-i\zeta_3\sigma_2\partial_\nu+\zeta_3\sigma_3\bigl[k(\tau_1\sinh\mu+\tau_2\sin\nu)-q\bigr]\\
&+j\zeta_1(\tau_1\csc\nu+\tau_2\operatorname{csch}\mu).
\end{split}
\end{equation}
The superscript $(3)$ indicates that $H^{(3)}_{k,q,j}$ arises from the three-dimensional Dirac operator, while $j$ labels the
axial angular-momentum sector. In particular, $H^{(3)}_{k,q,j}$ depends on $q$ through a bounded perturbation
\begin{equation}\label{ax:bounded-perturbation}
 H^{(3)}_{k,q,j}=H^{(3)}_{k,0,j}-qJ,
 \quad \text{with } J:=\zeta_3\sigma_3 \text{ satisfying that } J^2=I.
\end{equation}
The reason we introduce this axial operator is twofold: 1) thanks to its explicit expression, its spectrum can be estimated by exact polynomial inequalities, which consequently produces the estimate of the two-center Birman-Schwinger operator (see Proposition \ref{ax:singular-vector} below). We will illustrate the intuition behind these technical exact numeric inequalities whenever they appear and leave the technical details to the appendices.

For completeness of the argument, the basic properties of the auxiliary axial operator are summarized as follows. As the proof is technical and does not contribute to the major goal of this paper, we put it in Appendix~\ref{sec:H_kqj}.

\begin{proposition}[Characterization of $H^{(3)}_{k,q,j}$]
\label{ax:domain}
For every $k>0$ and $j\in\mathbb Z+\tfrac12$, the operator $H^{(3)}_{k,q,j}$ is essentially self-adjoint on $C_c^\infty(\Omega;\mathbb C^8)$. Its closure has compact resolvent.  For fixed $k,j$, the domain $\Dom(H^{(3)}_{k,q,j})$ is independent of $q$, and \(C_c^\infty(\Omega;\mathbb C^8)\) is a graph core for every \(q\in\mathbb R\).
\end{proposition}
In the sequel, we denote the unique self-adjoint closure of $H^{(3)}_{k,q,j}$ in $L^2(\Omega;\mathbb C^8)$ by the same symbol. The next proposition justifies the claim that the spectrum of the auxiliary operator determines the norm of the two-center Birman-Schwinger operator:

\begin{proposition}[The singular-value correspondence]
\label{ax:singular-vector}
Let $0\leq E<m$, $\kappa:=(m^2-E^2)^{1/2}$ and $k=a\kappa$. The operator
\[
 T_E=w_B(D_m-E)^{-1}w_A
\]
is compact and commutes with the axial angular momentum operator $J_{z}$ on $\Dom(J_{z})$. If $s>0$ is a singular value of $T_E$ in the sector $j\in\mathbb Z+\tfrac12$ (of $J_z$), then $H^{(3)}_{k,q,j}$ has a nonzero eigenvector satisfying
\begin{equation}\label{ax:physical-ray}
 H^{(3)}_{k,q,j}\Phi=\lambda\Phi,
 \quad \text{with } \lambda=\frac{m}{s\kappa} \,\text{ and }\, q=\frac{E}{m}\lambda.
\end{equation}
\end{proposition}

By Proposition \ref{ax:singular-vector}, in order to prove the two-center estimate in Theorem \ref{thm:cross}, it therefore suffices to exclude $[-1,1]$ from the spectrum of $H^{(3)}_{k,q,j}$ for every $k>0$, $q\in[0,1]$, and $j\in\mathbb Z+\tfrac12$. This will be the main focus of Sections \ref{end:section}-\ref{sec:bridge}, where we will prove that $[-1,1]\cap \text{spec}(H^{(3)}_{k,q,j}) = \varnothing$ in different regimes of $k,q$.

In the sequel, Section~\ref{sec:T_E} proves Proposition~\ref{ax:singular-vector} by a change of variables and spinor transformations. These transformations lead to the transfer of singular values as stated in Proposition~\ref{ax:singular-vector}, which is the key point that justifies the purpose of introducing the auxiliary operator.

\subsection{Proof of Proposition \ref{ax:singular-vector}} \label{sec:T_E}

\subsubsection{Compactness and commutation} \label{sec:compact_commute}

We first show the compactness of $T_{E}$. First, it is elementary that the multiplication by $w_B=|x-B|^{-1/2}$ is compact from $H^1(\mathbb{R}^3;\mathbb{C}^4)$ to $L^2(\mathbb{R}^3;\mathbb{C}^4)$; consequently, its composition with the resolvent, i.e., $w_B(D_{m}-E)^{-1}$, is compact on $L^2(\mathbb{R}^3;\mathbb{C}^4)$. However, since $w_{A}$ is not bounded as a multiplication operator on $L^2$, we should be careful and do not directly claim that the second composition $T_{E}=w_B(D_{m}-E)^{-1}w_{A}$ is compact. This is resolved by a standard truncation argument. More precisely, we select disjoint neighborhoods of $A$ and $B$, choose smooth compactly supported cutoffs $\chi_A,\chi_B$ in these neighborhoods, each equal to one near its center, and write $w_{X}^{int}:=\chi_{X}w_{X}$, $w_{X}^{ext}:=(1-\chi_{X})w_{X}$ for $X\in \{A,B\}$. Then $T_{E}$ decomposes as
\begin{equation*}
T_{E}=w_B(D_{m}-E)^{-1}w_{A}^{ext}+w_B^{ext}(D_{m}-E)^{-1}w_{A}^{int}+w_B^{int}(D_{m}-E)^{-1}w_{A}^{int}
\end{equation*}
Note that
\begin{itemize}
    \item $w_B(D_{m}-E)^{-1}w_{A}^{ext}$ is compact as $w_{A}^{ext}$ is $L^{\infty}$ bounded and is therefore bounded as a multiplication operator on $L^2$;
    \item $w_B^{ext}(D_{m}-E)^{-1}w_{A}^{int}$ is compact because 1) $(D_{m}-E)^{-1}w_{A}^{int}$ is compact as the adjoint of the compact operator $w_A^{int}(D_m-E)^{-1}$, and 2) $w_{B}^{ext}$ is $L^{\infty}$ bounded;
    \item $w_B^{int}(D_{m}-E)^{-1}w_{A}^{int}$ is Hilbert-Schmidt by investigating its integral kernel $w_B^{int}(x)(D_{m}-E)^{-1}(x,y)w_{A}^{int}$ and noting that 1) $(D_{m}-E)^{-1}(x,y)$ is bounded as $x,y$ are separated, and 2) $w_B^{int}(\cdot),w_A^{int}(\cdot)\in L^2$.
\end{itemize}
This proves the compactness of $T_E$.

Now, we prove the commutation between $T_{E}$ and $J_{z}$ by direct calculation. Denoting the momentum operator $\bm{p}:=-i\nabla$, we have $D_{m}=\bm{\alpha}\cdot \bm{p}+m\beta$ and $L_{z}=xp_{y}-yp_{x}$. Then, using the formal commutation relations
\[
[L_z,p_x]=ip_y,\quad[L_z,p_y]=-ip_x,\quad [L_z,p_z]=0,
\]
we can then check that
\begin{equation} \label{eq:TE_Jz_commutation_1}
[L_z,D_m]=i\alpha_xp_y-i\alpha_yp_x .
\end{equation}
Moreover, by the Clifford identities, it holds that
\[
[S_z,\alpha_x]=i\alpha_y,\quad [S_z,\alpha_y]=-i\alpha_x,\quad [S_z,\alpha_z]=[S_z,\beta]=0
\]
by which we know that
\begin{equation} \label{eq:TE_Jz_commutation_2}
[S_z,D_m]=i\alpha_yp_x-i\alpha_xp_y.
\end{equation}
By \eqref{eq:TE_Jz_commutation_1} and \eqref{eq:TE_Jz_commutation_2}, we know $[J_z,D_m]=0$ on \(C_c^\infty(\mathbb R^3;\mathbb C^4)\), by which the commutation claim on the free resolvent follows using the unitary group generated by $J_{z}$. Finally, noting that the scalar potential weights $w_A$ and $w_B$ are independent of the variable $\varphi$, and hence commute with both $L_{z}$ and $S_{z}$, we conclude the proof of commutation between $T_{E}$ and $J_{z}$.

\subsubsection{Transfer of singular values} \label{sec:singular_transfer}

We now prove the remaining claim in Proposition \ref{ax:singular-vector}, which states that a singular value of $T_{E}$ transfers to an eigenvalue of $H_{k,q,j}^{(3)}$. This is achieved by constructing a sequence of consecutive transformations that transform the operator $T_{E}$ to $H_{k,q,j}^{(3)}$. To construct these transformations explicitly, we fix the Clifford representation of $\alpha_i$ and $\beta$ throughout this section:
\begin{equation} \label{ax:clifford}
 \alpha_z=\sigma_1,\quad \alpha_x=\sigma_2, \quad \alpha_y=\sigma_3\zeta_1, \quad \beta=\sigma_3\zeta_3.
\end{equation}
Recall that $\Sigma_z=-i\alpha_x\alpha_y=\sigma_1\zeta_1$ is the (axial) spin generator about the $z$-axis. In the sequel, for readers' convenience, we will state explicitly in the subheading of each step the domain/codomain of each transformation that is constructed therein.

{\color{blue}Step 1. Restriction to the $j$\textsuperscript{th} angular sector: $L^2(Q;\mathbb C^4)\to\mathcal H_j$.}

First, in cylindrical coordinates $(z,\rho,\varphi)$, we introduce the unitary map $\mathscr S_j:L^2(Q;\mathbb C^4)\to\mathcal H_j$, mapping the meridional space to the $j$\textsuperscript{th} sector $\mathcal H_j=\ker(J_z-j)$:
\begin{equation}\label{ax:sector-map}
(\mathscr S_j\psi)(z,\rho,\varphi):=(2\pi\rho)^{-1/2}
   e^{-i\varphi\Sigma_z/2}e^{ij\varphi}\psi(z,\rho).
\end{equation}
The half-integer values of $j$ make this spinor single-valued; hence, by direct differentiation, one checks that $J_z\mathscr S_j\psi=j\mathscr S_j\psi$. Moreover, $\mathscr S_j$ is unitary, as checked by noticing the identity
\[
 \|\mathscr S_j\psi\|_{L^2(\mathbb R^3)}^2 =\frac1{2\pi}\int_Q\int_0^{2\pi}|\psi(z,\rho)|^2       \,d\varphi\,dz\,d\rho
 =\|\psi\|_{L^2(Q)}^2.
\]
With this notion, the restriction of the free Dirac operator on the $j$\textsuperscript{th} sector is defined as
\begin{equation}\label{ax:meridional}
D_{m,j}:= \mathscr S_j^*D_m\mathscr S_j=-i\sigma_1\partial_z-i\sigma_2\partial_\rho
       +\sigma_3\left(m\zeta_3+\frac j\rho\zeta_1\right).
\end{equation}
Suppose $s>0$ is a singular value of $T_{E}$ in the $j$\textsuperscript{th} sector, i.e., for some spinors $f,h\in \mathcal H_j$,
\begin{equation} \label{eq:singular_doublet}
T_Ef=sh,\quad T_E^*h=sf.
\end{equation}
We define
\begin{equation} \label{eq:tilde_hf}
\widetilde h:=\mathscr S_j^*h,\quad
 \widetilde f:=\mathscr S_j^*f,\quad
 \psi:=\sqrt{r_B}\,\widetilde h,\quad
 \chi:=\sqrt{r_A}\,\widetilde f.
\end{equation}
Then the spinors $\psi$ and $\chi$ satisfy the following equations with $g:=s^{-1}$
\begin{equation}\label{ax:coupled}
(D_{m,j}-E)\sqrt{r_B}\,\widetilde h=\frac g{\sqrt{r_A}}\,\widetilde f, \quad
 (D_{m,j}-E)\sqrt{r_A}\,\widetilde f=\frac g{\sqrt{r_B}}\,\widetilde h.
\end{equation}

{\color{blue}Step 2. Prolate transform: $L^2(Q;\mathbb C^4)\to
 L^2(\Omega;\mathbb C^4)$.}
 
Next, we introduce the conformal coordinate map $\Xi_a:(\mu,\nu)\in \Omega \mapsto (z,\rho)\in Q$ defined by
\begin{equation}\label{ax:coordinates}
 z+i\rho=a\cosh(\mu+i\nu) .
\end{equation}
By this transform, we are able to treat the two centers of potential, $A$ and $B$, symmetrically:
\begin{equation*}
r_A=a(\cosh\mu+\cos\nu),\quad r_B=a(\cosh\mu-\cos\nu).
\end{equation*}
In particular, exchanging the centers corresponds simply to \(\nu\mapsto\pi-\nu\). Moreover, the two distances enter through a common conformal factor as seen by
\[
dz^2+d\rho^2 =r_A r_B\,(d\mu^2+d\nu^2).
\]
Accordingly, the pullback associated with the transform \eqref{ax:coordinates} is denoted as
\begin{equation}\label{ax:coordinate-unitary}
 \mathscr C_a:L^2(Q;\mathbb C^4)\to
 L^2(\Omega;\mathbb C^4),\quad
 (\mathscr C_a v)(\mu,\nu)
 :=\sqrt{r_A r_B}\,v(\Xi_a(\mu,\nu)),
\end{equation}
which is also unitary by noting that $\|\mathscr C_a v\|_{L^2(\Omega)}^2=\int_Q|v|^2\,dz\,d\rho$.

We record the image of the doubled map $\mathscr C_a\oplus\mathscr C_a$ applied on the pair $(\widetilde h,\widetilde f)$ in \eqref{eq:tilde_hf} by
\begin{equation}\label{ax:coordinate-pair}
\xi:=(\mathscr C_a\oplus\mathscr C_a)(\widetilde h,\widetilde f).
\end{equation}

{\color{blue}Step 3. Scaled permutation: $L^2(\Omega;\mathbb C^8)\to
 L^2(\Omega;\mathbb C^8)$.}

Now we introduce the non-vanishing scaling factors for $(\mu,\nu)\in \Omega$:
\[
\alpha:=\sqrt{2a}\cosh\frac{\mu+i\nu}{2},\quad
\gamma:=\sqrt{2a}\sinh\frac{\mu+i\nu}{2}.
\]
In particular, they satisfy the identities
\begin{equation}\label{eq:alpha_gamma}
\begin{gathered}
\alpha^2=z+i\rho+a,\quad \gamma^2=z+i\rho-a,\quad
|\alpha|^2=r_A,\quad |\gamma|^2=r_B,\\
\alpha\overline\gamma=a(\sinh\mu-i\sin\nu).
\end{gathered}
\end{equation}
With the factors $\alpha,\gamma$, we define the scaled permutation on
$L^2(\Omega;\mathbb C^8)$:
\begin{equation}\label{ax:root-unitary}
\mathscr R_a
\begin{pmatrix}\xi_1\\\xi_2\\\xi_3\\\xi_4\end{pmatrix}
:=\begin{pmatrix}
   (\alpha/\sqrt{r_A})\xi_1\\
   (\gamma/\sqrt{r_B})\xi_3\\
   (\overline\gamma/\sqrt{r_B})\xi_4\\
   (\overline\alpha/\sqrt{r_A})\xi_2
  \end{pmatrix}.
\end{equation}
Since the multiplier in each entry has modulus one as checked using \eqref{eq:alpha_gamma}, it follows that $\mathscr R_a$ is unitary. As before, we record the image
\begin{equation*}
(U,W)^\top:=\mathscr R_a\xi .
\end{equation*}
Then one can check that equations~\eqref{ax:coupled}
become
\begin{equation}\label{ax:unbalanced}
\mathcal D_+U=i\{g+(M+E)Z\}W,\quad
\mathcal D_-W=i\{g-(M-E)Z\}U ,
\end{equation}
with 
\begin{equation*}
\mathcal D_\pm:=\partial_\mu\pm i\partial_\nu,\quad
M:=m\zeta_3+j\zeta_1/\rho ,
\end{equation*}
and
\begin{equation*}
Z:=a(\tau_1\sinh\mu+\tau_2\sin\nu)
\end{equation*}
which acts on the \(\tau\)-components and is understood as identity on the \(\zeta\)-components.

{\color{blue}Step 4. Mass balancing: $L^2(\Omega;\mathbb C^8)\to L^2(\Omega;\mathbb C^8)$.}

Finally, we introduce another scaling to balance the two meridional mass signs in \eqref{ax:unbalanced}: defining the constant positive matrix
\begin{equation}\label{ax:mass-balance}
 P:=\left(\frac{m+E\zeta_3}{\kappa}\right)^{1/2},
\end{equation}
the balancing map is defined as
\begin{equation}\label{ax:balancing-map}
\mathscr B_E\binom UW:=\binom{P^{-1}U}{P\zeta_3W} =:\binom FL.
\end{equation}
Note that $\mathscr B_E$ is continuous and has a bounded inverse for $E\in [0,m)$. In particular, when $E=0$, we know $P=I$ and hence $\mathscr B_0(U,W)=(U,\zeta_3W)$ is unitary by noting that $\zeta_3^*\zeta_3=I$. For $E>0$, $\mathscr B_E$ is not unitary, as seen by
\[
 \|\mathscr B_E(U,W)^{\top}\|^2
 =\|P^{-1}U\|^2+\|PW\|^2.
\]
Now we see how the singular value of $T_{E}$ transfers: since $\det P=1$, it can be checked that
\[
 P^{-1}\zeta_1\zeta_3P^{-1}=\zeta_1\zeta_3,
 \qquad P\zeta_3\zeta_1P=\zeta_3\zeta_1.
\]
Consequently, with $\mathcal J:=(j/\rho)\zeta_1\zeta_3Z$, $\lambda:=gm/\kappa$ and $q:=gE/\kappa$, we obtain
\begin{equation}\label{ax:balanced}
\begin{split}
 \mathcal D_+F&=i\{\lambda\zeta_3-q+\kappa Z+\mathcal J\}L,\\
 \mathcal D_-L&=i\{\lambda\zeta_3+q-\kappa Z+\mathcal J\}F.
\end{split}
\end{equation}
For $\Phi=(F,L)^\top$, multiplying \eqref{ax:balanced} by $\zeta_3$ exactly establishes the desired identities \eqref{ax:physical-ray} in the interior of $\Omega$. To conclude that $\Phi\in \Dom(H_{k,q,j}^{(3)})$, we first claim that $\Phi\in \Dom(H_{k,q,j,int}^{(3),*})$, where $H_{k,q,j,int}^{(3)}$ denotes the formal operator \eqref{ax:pencil} restricted to $C_c^\infty(\Omega;\mathbb C^8)$, and $H_{k,q,j,int}^{(3),*}$ denotes its adjoint; then the fact $\Phi\in \Dom(H_{k,q,j}^{(3)})$ is justified by the essential self-adjointness of $H_{k,q,j}^{(3)}$, which is proved in Appendix~\ref{sec:H_kqj} using the transformation constructed here. Since \eqref{ax:pencil} is formally symmetric with smooth coefficients, the adjoint $H_{k,q,j,int}^{(3),*}$ has the maximal distributional domain
\begin{equation}\label{ax:maximal-domain}
 \Dom(H_{k,q,j,int}^{(3),*}) =\bigl\{\Theta\in L^2(\Omega;\mathbb C^8):
       H^{(3)}_{k,q,j}\Theta\in L^2(\Omega;\mathbb C^8)
       \text{ in }\mathcal D^{\prime}(\Omega;\mathbb C^8)\bigr\}.
\end{equation}
Hence, by testing equation \eqref{ax:physical-ray} with $\Theta\in C_{c}^{\infty}(\Omega;\mathbb C^8)$, one sees that $\Phi\in \Dom(H_{k,q,j,int}^{(3),*})$ and $H_{k,q,j,int}^{(3),*}\Phi=\lambda \Phi$. This completes the proof of Proposition \ref{ax:singular-vector}.

{\color{blue}Step 5. Summary: $\mathcal H_j\oplus\mathcal H_j\to
 L^2(\Omega;\mathbb C^8)$.}

Now, we record the transformation we have constructed for later use: 
\begin{equation}\label{ax:unitary-composition}
 \mathcal U_{j,E}:\mathcal H_j\oplus\mathcal H_j
 \longrightarrow L^2(\Omega;\mathbb C^8),\quad
 \mathcal U_{j,E}
 :=\mathscr B_{E}\mathscr R_a
   (\mathscr C_a\oplus\mathscr C_a)
   (\mathscr S_j^*\oplus\mathscr S_j^*).
\end{equation}
For $E=0$, each factor in $\mathcal U_{j,0}$ is onto and norm preserving, i.e., $\|\mathcal U_{j,0}(h,f)\|^2=\|h\|^2+\|f\|^2$. For $E\neq 0$, $\mathcal U_{j,E}$ is not unitary.

\subsection{The zero-energy inverse}\label{sec:axial-inverse}

For later use, let $k=am$ and let $\mathcal U_{j,0}$ be the unitary transformation in \eqref{ax:unitary-composition} at $E=0$. On the sector $\mathcal H_j\oplus\mathcal H_j$, we write the block Birman-Schwinger operator at zero energy and its transformation as
\begin{equation}\label{ax:doubled-resolvent}
 C_0:=\begin{pmatrix}0&T_0\\T_0^*&0\end{pmatrix},
 \quad K:=\mathcal U_{j,0}C_0\mathcal U_{j,0}^* .
\end{equation}
With these notions, the transformation constructed in the last subsection identifies the inverse of the auxiliary operator $H^{(3)}_{k,0,j}$ with K:
\begin{equation}\label{ax:inverse-identity}
(H^{(3)}_{k,0,j})^{-1}=K.
\end{equation}
For completeness, a detailed proof of this identity is given in Appendix~\ref{sec:H_kqj}. The key ingredient needed to study the spectrum of $H^{(3)}_{k,q,j}$ in the large-$k$ (i.e., large separation of the two centers $A$ and $B$) regime in later sections is the estimate of the norm of zero-energy inverse $(H^{(3)}_{k,0,j})^{-1}$, which is presented in the subsequent paragraphs.

\subsection{The zero-energy gap}

\begin{proposition}\label{ax:zero-gap}
For $k>0$ and $j\in\mathbb Z+\tfrac12$,
\begin{equation}\label{ax:zero-gap-bound}
 \bigl(H^{(3)}_{k,0,j}\bigr)^2
 \geq \left(|j|+\tfrac12\right)^2
       +k\left(2|j|+\frac{1}{2(|j|+1)}\right)
 \geq1+\frac{4k}{3}.
\end{equation}
The inequalities are understood in the sense of the associated quadratic form on $\Dom(H^{(3)}_{k,0,j})$.
\end{proposition}

\begin{proof}

{\color{blue}Step 1. Preparations.}

Since conjugation by the unitary matrix $\zeta_3$ changes $H^{(3)}_{k,0,j}$ into $H^{(3)}_{k,0,-j}$ and preserves $C_c^\infty(\Omega;\mathbb C^8)$, it suffices to prove the estimate for $j>0$. In the sequel, we first prove the estimate on $C_c^\infty(\Omega;\mathbb C^8)$ and then extend to the full domain $\Dom(H^{(3)}_{k,0,j})$. To calculate the squared operator, we write
\[
X:=\tau_1\sinh\mu+\tau_2\sin\nu,
\quad
Y:=\tau_1\csc\nu+\tau_2\operatorname{csch}\mu,
\]
by which it follows that
\begin{equation*}
H^{(3)}_{k,q,j}=-i\zeta_3\sigma_1\partial_\mu-i\zeta_3\sigma_2\partial_\nu+\zeta_3\sigma_3\big(kX-q\big)
+j\zeta_1 Y
\end{equation*}
It is direct to check that
\begin{equation}\label{ax:commutator}
[X,Y]=2i\tau_3\bigl(\sinh\mu\operatorname{csch}\mu-\sin\nu\csc\nu\bigr)=0,
\end{equation}
using the commutation relation of Pauli matrices. Remarkably, this exact commutation $[X,Y]=0$ implies that, when calculating the square $\bigl(H^{(3)}_{k,0,j}\bigr)^2$, the mixed terms containing the factor $jk$ cancel out:
\begin{equation}\label{ax:separated-square}
\begin{split}
\bigl(H^{(3)}_{k,0,j}\bigr)^2
={}&-\partial_\mu^2-\partial_\nu^2
+k^2(\sinh^2\mu+\sin^2\nu)\\
&-k\sigma_2\tau_1\cosh\mu+k\sigma_1\tau_2\cos\nu
+j^2(\operatorname{csch}^2\mu+\csc^2\nu)\\
&-j\zeta_2\bigl(\sigma_1\tau_2\operatorname{csch}\mu\coth\mu
+\sigma_2\tau_1\csc\nu\cot\nu\bigr).
\end{split}
\end{equation}
Moreover, we point out that this squared operator has a convenient structure that \textit{the dependence on the variables $\mu$ and $\nu$ are separated exactly}, which provides a shortcut to estimate the lower bound of $\bigl(H^{(3)}_{k,0,j}\bigr)^2$. To see this, we first note that the three Hermitian operators $\sigma_2\tau_1$, $\sigma_1\tau_2$ and $\zeta_2$ commute. We may therefore decompose the spinor space into their simultaneous eigenspaces. On an eigenspace with respective eigenvalues $a_0,b_0,c_0\in\{-1,1\}$, the squared operator \eqref{ax:separated-square} separates into
\begin{equation} \label{ax:separated-square_1}
\bigl(H^{(3)}_{k,0,j}\bigr)^2=R_{a_0,p}+A_{b_0,\varepsilon}
\end{equation}
where $R_{a_0,p}$ and $A_{b_0,\varepsilon}$ depend only on $\mu$ and $\nu$, respectively,
\begin{align}
R_{a_0,p}
&:=-\partial_\mu^2+k^2\sinh^2\mu-a_0k\cosh\mu
+j^2\operatorname{csch}^2\mu
-pj\operatorname{csch}\mu\coth\mu,
\label{ax:radial-forms}\\
A_{b_0,\varepsilon}
&:=-\partial_\nu^2+k^2\sin^2\nu+b_0k\cos\nu
+j^2\csc^2\nu-\varepsilon j\csc\nu\cot\nu,
\label{ax:angular-forms}
\end{align}
and $p=c_0b_0$, $\varepsilon=c_0a_0$. 

{\color{blue}Step 2. Individual estimates.}

It remains to estimate the radial operator \eqref{ax:radial-forms} and the angular operator \eqref{ax:angular-forms} while keeping track of the relation between their signs. As one-dimensional operators, both the lower bounds of $R_{a_0,p}$ and $A_{b_0,\varepsilon}$ are obtained using the following elementary identity: for a one-dimensional (formal) differential operator $\mathcal L=-\partial_x^2+V$ on a (possibly non-compact) interval $I$, any positive smooth function $f$, and $u\in C_c^\infty(I)$, integration by parts gives
\begin{equation} \label{eq:ground_state_estimate}
\int_I\bigl(|u'|^2+V|u|^2\bigr)\,dx
=
\int_I f^2\left|\left(\frac{u}{f}\right)'\right|^2\,dx
+\int_I\frac{\mathcal Lf}{f}|u|^2\,dx.
\end{equation}
In particular, a pointwise lower bound $\mathcal Lf/f\geq c$ implies that $\mathcal L\geq c$, which holds in the sense of quadratic form on compactly supported test functions.

{\color{blue}Step 2.1. Estimate of the radial part.} We now estimate the radial operator $R_{a_0,p}$ defined in \eqref{ax:radial-forms} on the four sectors indexed by $a_0,p\in \{\pm 1\}$, respectively. To this end, we define
\[
Q:=\partial_\mu+k\sinh\mu-j\operatorname{csch}\mu.
\]
Then, denoting $Q^{*}$ by its formal adjoint, we know
\[
R_{1,1}=Q^*Q+2kj\geq2kj.
\]
For $R_{1,-1}$, we choose the positive function
\[
f_-(\mu):=e^{-k\cosh\mu}\tanh(\mu/2)^{j+1},
\]
for which one can verify that
\[
\frac{R_{1,-1}f_-}{f_-}=2k(j+1)+\frac{2j+1}{\cosh\mu+1}\geq 2k(j+1).
\]
Then, using \eqref{eq:ground_state_estimate}, we know $R_{1,-1}\geq2k(j+1)$. The other sectors are treated similarly. In summary,
\begin{equation} \label{ax:radial-bounds}
R_{1,1}\geq 2kj,\quad R_{1,-1}\geq 2k(j+1),\quad R_{-1,1}\geq 2k(j+1),\quad R_{-1,-1}\geq 2k(j+2).
\end{equation}

{\color{blue}Step 2.2. Estimate of the angular part.} We next estimate the angular operator defined in \eqref{ax:angular-forms} based on similar arguments as in Step 2.1. Let $n:=j+\tfrac12$ and choose
\[
f_0(\nu):=(\sin\nu)^n\tan(\nu/2)^{-\varepsilon/2},
\quad
t_0:=\frac{kb_0}{2(j+1)},
\quad
f(\nu):=e^{-t_0\cos\nu}f_0(\nu).
\]
The two functions $f$ and $f_0$ are both positive on $0<\nu<\pi$. In particular, one can check that
\[
-\frac{f_0''}{f_0}
+j^2\csc^2\nu-\varepsilon j\csc\nu\cot\nu=n^2.
\]
Hence, using $f'/f=t_0\sin\nu+f_0'/f_0$, we obtain
\[
\begin{aligned}
\frac{A_{b_0,\varepsilon}f}{f}
&=n^2+\varepsilon t_0+(k^2-t_0^2)\sin^2\nu
+\bigl(kb_0-2(j+1)t_0\bigr)\cos\nu\\
&=n^2+\varepsilon t_0+(k^2-t_0^2)\sin^2\nu.
\end{aligned}
\]
Since $|t_0|<k$, the identity \eqref{eq:ground_state_estimate} gives
\begin{equation}\label{ax:angular-bound}
A_{b_0,\varepsilon}\geq
\left(j+\tfrac12\right)^2
+\frac{kb_0\varepsilon}{2(j+1)}
=\left(j+\tfrac12\right)^2
+\frac{ka_0 p}{2(j+1)},
\end{equation}
where the last equality follows by recalling the relation $b_0\varepsilon=a_0p$ proved by the definitions following \eqref{ax:angular-forms}.

{\color{blue}Step 2.3. Combination.} Finally, we combine the results from the previous two steps. Using \eqref{ax:radial-bounds}, \eqref{ax:angular-bound} and the decomposition $\bigl(H^{(3)}_{k,0,j}\bigr)^2=R_{a_0,p}+A_{b_0,\varepsilon}$, the lower bound of $\bigl(H^{(3)}_{k,0,j}\bigr)^2$ is obtained as follows corresponding the four choices of the radial index $(a_0,p)$:
\[
R_{a_0,p}+A_{b_0,\varepsilon}-\left(j+\tfrac12\right)^2
\geq k
\begin{cases}
2j+\dfrac{1}{2(j+1)},
& (a_0,p)=(1,1),\\[4pt]
2(j+1)-\dfrac{1}{2(j+1)},
& (a_0,p)=(1,-1)\text{ or }(-1,1),\\[4pt]
2(j+2)+\dfrac{1}{2(j+1)},
& (a_0,p)=(-1,-1).
\end{cases}
\]
Since $j\geq\tfrac12$, the smallest coefficient is achieved when $(a_0,p)=(+1,+1)$. This proves the desired inequality \eqref{ax:zero-gap-bound} on $C_c^\infty(\Omega;\mathbb C^8)$. The extension to the full domain $\Dom(H_{k,0,j}^{(3)})$ is guaranteed by the fact that $C_c^\infty(\Omega;\mathbb C^8)\subset \Dom(H_{k,0,j}^{(3)})$ forms a graph core, as stated in Proposition~\ref{ax:domain}. The details are omitted here.
\end{proof}

\begin{corollary}\label{ax:zero-cross}
For distinct $A,B\in\mathbb R^3$ and $m>0$,
\begin{equation}\label{ax:zero-cross-bound}
 \|w_BD_m^{-1}w_A\|
 \leq\left(1+\frac{2m|A-B|}{3}\right)^{-1/2}<1.
\end{equation}
\end{corollary}
\begin{proof}
For $E=0$, the singular-value correspondence is $\lambda=s^{-1}$ and $k=m|A-B|/2$. Then the estimate of singular values of $w_BD_m^{-1}w_A$ at each $j$-sector, following by Proposition \ref{ax:zero-gap}, yields \eqref{ax:zero-cross-bound}.
\end{proof}

For later use, we define the gap function
\begin{equation}\label{ax:gap-function}
 \nu_{k,j}(q):=\inf_{\substack{ \Phi\in\Dom(H^{(3)}_{k,0,j})\\\|\Phi\|=1}}
                         \|H^{(3)}_{k,q,j}\Phi\|^2,
\end{equation}
which is well-defined since $H^{(3)}_{k,0,j}$ has a compact resolvent. In particular, $\nu_{k,j}(q)$ is continuous in $q$:
\begin{equation}\label{ax:gap-continuity}
 |\sqrt{\nu_{k,j}(q)}-\sqrt{\nu_{k,j}(q')}|
 \leq|q-q'|,
\end{equation}
which follows by the estimate $\|H_q\Phi\| \le \|H_{q'}\Phi\|+|q-q'|$ as a consequence of the decomposition \eqref{ax:bounded-perturbation}. Moreover, $q\mapsto\nu_{k,j}(q)-q^2$ is concave, as the right side is the infimum over unit vectors of the affine functions
\[
 \|H^{(3)}_{k,0,j}\Phi\|^2
 -2q\operatorname{Re}\langle H^{(3)}_{k,0,j}\Phi,J\Phi\rangle.
\]
In particular, for $|j|\geq3/2$ and $0\leq q\leq1$, Proposition \ref{ax:zero-gap} implies that
\begin{equation}\label{ax:higher-sectors}
 \sqrt{\nu_{k,j}(q)}
 \geq\sqrt{\left(|j|+\tfrac12\right)^2+
             k\left(2|j|+\frac1{2(|j|+1)}\right)}-q>1.
\end{equation}
In the sequel, we will need to show that $\sqrt{\nu_{k,j}(q)}>1$ for all $j$; thanks to \eqref{ax:higher-sectors}, we only need to consider the sectors $|j|=\frac{1}{2}$.

\section{Large-separation extension of the gap estimate}
\label{end:section}

In this section, we extend the zero-energy gap obtained in Proposition \ref{ax:zero-gap} to a range of nonzero values of $q$. The main result of this section is Corollary \ref{end:large-k}, which extends the gap estimate to the parameter range
\begin{equation*}
k\geq 2/5,\quad 0\leq q\leq 1.
\end{equation*}
This implies that, in the above parameter range (the large-separation regime), the interval $[-1,1]$ is excluded from the spectrum of $H^{(3)}_{k,q,j}$. This will serve as the first part for the proof of Theorem \ref{thm:cross} using Proposition \ref{ax:singular-vector}.

The main observation to extend the gap condition to the specified range is as follows. First, recall that $H^{(3)}_{k,q,j}$ differs from $H^{(3)}_{k,0,j}$ by a bounded perturbation $qJ$; see \eqref{ax:bounded-perturbation}. Importantly, compressing the inverse $(H^{(3)}_{k,0,j})^{-1}$ to the two eigenspaces of $J$ reduces, through the unitary transformation constructed in Section \ref{sec:singular_transfer}, to an estimate for the scalar resolvent $(-\Delta+m^2)^{-1}$ with two Coulomb weights; see \eqref{eq:compressed_inverse}. We first prove this scalar estimate (in Lemma \ref{end:scalar-cross}) and then use it to obtain an endpoint bound for the gap function (in Proposition \ref{end:endpoint}). Finally, concavity gives the estimate between zero and this endpoint; see Corollary \ref{end:large-k}.

\begin{lemma} \label{end:scalar-cross}
Let $\kappa>0$ and $A,B\in\mathbb R^3$ be distinct. Let $d:=\kappa|A-B|$,
\begin{equation}\label{end:overlap-functions}
s(d):=(1+d)e^{-d},
\quad
c(d):=\frac{\sqrt{8+s(d)^2}+s(d)}4,
\end{equation}
and recall that $w_X:=|x-X|^{-1/2}$ ($X\in \{A,B\}$). Then $1/\sqrt2<c(d)<1$ and
\begin{equation}\label{end:scalar-bound}
\big\|w_{B}(-\Delta+\kappa^2)^{-1}w_{A}\big\|
\leq\frac{c(d)}{2\kappa}.
\end{equation}
\end{lemma}

\begin{proof}

By the unitary dilation map $y=\kappa x$, it is clear that we only need to prove the estimate \eqref{end:scalar-bound} for $\kappa=1$, which is achieved by several steps as follows. Throughout this proof, the concerned operators act on the space $L^2(\mathbb R^3)$.

{\color{blue}Step 1. Reduction.}

For $X\in\{A,B\}$, we write
\[
L:=(-\Delta+1)^{-1},\quad
S_X:=w_XL^{1/2},\quad
P_X:=S_X^*S_X=L^{1/2}w_X^2L^{1/2}.
\]
By the one-weight compactness argument as in Section \ref{sec:compact_commute}, we know $S_X$ is compact, and hence $P_X$ is positive and compact. Then the operator norm in \eqref{end:scalar-bound} becomes $\|w_BLw_A\|=\|S_BS_A^*\|$, by which it is tempting to estimate the norm of $S_{X}$. However, we decompose for one more step using the polar decomposition of $S_{X}$:
\begin{equation} \label{eq:two_point_scalar}
\|w_BLw_A\|=\|S_BS_A^*\|=\|P_B^{1/2}P_A^{1/2}\|.
\end{equation}
As we shall see soon, the reason of doing this is that it is convenient to estimate the norm of $P_X$ using the known spectrum of the one-particle Coulomb Schrödinger operator. Indeed, an eigenvalue $\eta>0$ of $P_X$ corresponds, by setting $u:=L^{1/2}v$ for the associated eigenvector $v$, to
\[
\left(-\Delta-\frac{\eta^{-1}}{|x-X|}\right)u=-u.
\]
Since the negative Coulomb eigenvalues are $-1/(4\eta^2n^2)$ for $n\geq 1$ (cf. \cite[Theorem 10.9]{Teschl2009}), this implies $\eta=1/(2n)$ for some $n$. In particular, the largest eigenvalue of $P_X$ is $1/2$, which is simple and associated with the normalized eigenvector
\[
v_X:=(2\pi)^{-1/2}L^{-1/2}e^{-|x-X|}.
\]
Consequently, the positive operator is dominated by the first projection as follows
\begin{equation}\label{end:atomic-majorant}
P_X\leq\frac12Q_X,
\quad
Q_X:=\frac12I+\frac12|v_X\rangle\langle v_X|.
\end{equation}

To apply \eqref{eq:two_point_scalar} to obtain the estimate of $\|w_BLw_A\|$, we require one more ingredient than \eqref{end:atomic-majorant}, i.e., an overlapping integral characterizing the intersection between $Q_{A}$ and $Q_{B}$. This is given by
\begin{equation}\label{end:ground-overlap}
\langle v_B,v_A\rangle
=\frac1\pi\int_{\mathbb R^3}
             \frac{e^{-(r_A+r_B)}}{r_A}\,dx
=(1+d)e^{-d}=s(d)
\end{equation}
with $d:=|A-B|$. The calculation is elementary and hence omitted here.

{\color{blue}Step 2. The two-center estimate.}

Now we apply \eqref{eq:two_point_scalar}, \eqref{end:atomic-majorant} and \eqref{end:ground-overlap} to prove \eqref{end:scalar-bound}. First, we note that the right side of \eqref{eq:two_point_scalar} increases in each operator $P_{A}$ and $P_{B}$ in the sense that its value increases if one replaces $P_{X}$ by another operator that dominates it. Indeed,
\[
\|P_B^{1/2}P_A^{1/2}\|^2 =\|P_A^{1/2}P_BP_A^{1/2}\| =\|P_B^{1/2}P_AP_B^{1/2}\|,
\]
and the two expressions on the right side give the monotonicity in $P_B$ and $P_A$, respectively. Thus, \eqref{eq:two_point_scalar} and \eqref{end:atomic-majorant} imply
\[
\|w_BLw_A\|
\leq\frac12\|Q_B^{1/2}Q_A^{1/2}\|.
\]
It remains to calculate the norm on the right side. First, on the orthogonal complement of the two-dimensional space $\operatorname{span}\{v_A,v_B\}$, i.e., $\operatorname{span}\{v_A,v_B\}^{\perp}$, it is clear that $Q_B^{1/2}Q_A^{1/2}=\frac{1}{2}I$, which has operator norm $\frac{1}{2}$. On the other hand, $Q_B^{1/2}Q_A^{1/2}$ admits the following matrix form on $\operatorname{span}\{v_A,v_B\}$
\[
[Q_B^{1/2}Q_A^{1/2}]_{(v_A,v_B)}
=
\begin{pmatrix}
t&t(1-t)s\\
(1-t)s&t+(1-t)^2s^2
\end{pmatrix} :=M
\]
with $t=1/\sqrt{2}$ and $s:=\langle v_B,v_A\rangle$. Note that the basis $(v_A,v_B)$ is not orthonormal; its Gram matrix is given by
\[
G=\begin{pmatrix}1&s\\s&1\end{pmatrix}.
\]
Hence, the square of the norm of $Q_B^{1/2}Q_A^{1/2}$ on $\operatorname{span}\{v_A,v_B\}$ is given by the largest generalized eigenvalue of $M^*GMx=\lambda Gx$, which is calculated directly as
\[
\lambda_+ =\frac{4+s^2+s\sqrt{8+s^2}}8.
\]
Since $\lambda_+\geq 1/2$, which dominates the norm of $w_BLw_A$ on  $\operatorname{span}\{v_A,v_B\}^{\perp}$, we conclude that
\[
\|Q_B^{1/2}Q_A^{1/2}\|
=\sqrt{\lambda_+}
=\frac{\sqrt{8+s^2}+s}{4}.
\]
This completes the proof of \eqref{end:scalar-bound}.
\end{proof}

To apply the scalar estimate to the axial operator $H_{k,q,j}^{(3)}$, we introduce the constants
\begin{equation}\label{end:constants}
\begin{aligned}
&s_k:=s(2k)=(1+2k)e^{-2k},
\quad
c_k:=c(2k)=\frac{\sqrt{8+s_k^2}+s_k}{4},\quad
A_k:=1+\frac{4k}{3}.
\end{aligned}
\end{equation}
Recall the gap function $\nu_{k,j}$ defined in \eqref{ax:gap-function}. The next proposition gives an estimate of the gap function at $q=c_k^{-1}$ and then extends it to the interval $0\leq q\leq c_k^{-1}$ by concavity.

\begin{proposition}
\label{end:endpoint}
For every $k>0$ and $j\in\mathbb Z+\tfrac12$,
\begin{equation}\label{end:endpoint-bound}
\nu_{k,j}(c_k^{-1})\geq c_k^{-2}.
\end{equation}
Consequently, for $0\leq q\leq c_k^{-1}$,
\begin{equation}\label{end:chord}
\nu_{k,j}(q)\geq q^2+(1-c_kq)A_k.
\end{equation}
\end{proposition}

\begin{proof}

{\color{blue}Step 1. Estimates of the compressed inverse.}

Fix $a,m>0$ with $k=am$, and recall that $\mathcal U_{j,0}$ is the zero-energy unitary transformation in \eqref{ax:unitary-composition} and $J=\zeta_3\sigma_3$. For ease of notation, throughout the proof we write
\[
H_0:=H^{(3)}_{k,0,j},
\quad
H_q:=H^{(3)}_{k,q,j}=H_0-qJ.
\]
We first present a preliminary estimate on the compressed zero-energy resolvent $(I\pm J)^{1/2}H_0^{-1}(I\pm J)^{1/2}$:
\begin{equation}\label{end:compression}
\big\|(I\pm J)^{1/2}H_0^{-1}(I\pm J)^{1/2}\big\|\leq c_k.
\end{equation}
As we will see later in Step 2, this compressed resolvent arises naturally when considering the spectral gap of $H_{q}$. We now prove \eqref{end:compression} using the estimate in Lemma \ref{end:scalar-cross}. First, \eqref{ax:doubled-resolvent} gives
\[
H_0^{-1}=\mathcal U_{j,0}C_0\mathcal U_{j,0}^*.
\]
On the other hand, tracing the transformations defining $\mathcal U_{j,0}$ leads to
\[
\mathcal U_{j,0}^*J\mathcal U_{j,0}
=
\begin{pmatrix}
\beta&0\\
0&\beta
\end{pmatrix},
\]
where $\beta=\sigma_3\zeta_3$ is introduced in \eqref{ax:clifford}. Hence, noting the facts that 1) $\beta$ anticommutes with the kinetic Dirac matrices and 2) the identity $D_m^{-1}=D_m(-\Delta+m^2)^{-1}$, one then calculates that
\begin{equation} \label{eq:compressed_inverse}
(I\pm\beta) D_m^{-1} (I\pm\beta)
=\pm 2m (I\pm\beta) (-\Delta+m^2)^{-1}.
\end{equation}
Moreover, the scalar weights commute with $I\pm\beta$. Applying these identities to the two off-diagonal blocks of $C_0$ (recall the definition \eqref{ax:doubled-resolvent}) and noting the identity
\[
(I\pm\beta)^{1/2}=\frac{I\pm\beta}{\sqrt2},
\]
we obtain
\begin{equation*}
\big\|(I\pm J)^{1/2}H_0^{-1}(I\pm J)^{1/2}\big\|\leq
2m\|w_B(-\Delta+m^2)^{-1}w_A\|
\leq c_k,
\end{equation*}
where we have applied \(\|I\pm\beta\|=2\) to obtain the first inequality, and the second follows from Lemma~\ref{end:scalar-cross}, since $m|A-B|=2am=2k$. This concludes the proof of \eqref{end:compression}.

{\color{blue}Step 2. Exclusion of the end spectral points.}

We now use \eqref{end:compression} to show that
\begin{equation}\label{end:scalar-lines}
\{-q,q\}\cap\sigma(H_q)=\varnothing,
\qquad 0<q<c_k^{-1}.
\end{equation}
Suppose that $H_q\Phi=q\Phi$ for a nonzero $\Phi\in\operatorname{Dom}(H_q)$. Then, by the definition of $H_{q}$,
\[
H_0\Phi=q(I+J)\Phi.
\]
We claim that the vector $h:=(I+J)^{1/2}\Phi$ is nonzero: otherwise, $(I+J)\Phi=0$ and hence $H_0\Phi=0$, contradicting the invertibility of $H_0$. Consequently,
\[
(I+J)^{1/2}H_0^{-1}(I+J)^{1/2}h=q^{-1}h,
\]
which implies that $q^{-1}$ is an eigenvalue of the compressed inverse $(I+J)^{1/2}H_0^{-1}(I+J)^{1/2}$. This contradicts \eqref{end:compression} as we have imposed $q^{-1}>c_k$. The possible eigenvalue $-q$ is excluded by a similar argument after taking the negative sign in \eqref{end:compression}. On the other hand, since $H_q$ has compact resolvent by Proposition~\ref{ax:domain}, its spectrum consists of eigenvalues of finite multiplicity. This proves \eqref{end:scalar-lines}.

{\color{blue}Step 3. The gap estimates.}

Finally, we apply \eqref{end:scalar-lines} to prove Proposition \ref{end:endpoint}. Recall that the gap function $\nu_{k,j}$ is continuous by \eqref{ax:gap-continuity}, and Proposition~\ref{ax:zero-gap} gives
\[
\nu_{k,j}(0)\geq A_k>0.
\]
We claim that
\begin{equation} \label{eq:gap_function_end_point_proof_1}
\nu_{k,j}(q)>q^2,
\quad\text{for }0<q<c_k^{-1}.
\end{equation}
Otherwise, the continuity of $\nu_{k,j}$ implies that there exists $q_*\in(0,c_k^{-1})$ such that $\nu_{k,j}(q_*)=q_*^2$. Since $H_{q_*}$ has compact resolvent by Proposition \ref{ax:domain}, it holds that
\[
\nu_{k,j}(q_*)=\min_{\lambda\in\sigma(H_{q_*})}\lambda^2.
\]
Thus, either $q_*$ or $-q_*$ would be an eigenvalue of $H_{q_*}$, contradicting \eqref{end:scalar-lines}. This proves \eqref{eq:gap_function_end_point_proof_1}. Using \eqref{eq:gap_function_end_point_proof_1}, the continuity of $\nu_{k,j}$, and taking an approximating sequence that approaches $c_k^{-1}$, we conclude the proof of \eqref{end:endpoint-bound}.

Finally, recall that the function $g(q):=\nu_{k,j}(q)-q^2$ is concave, as proved after \eqref{ax:gap-continuity}. Since we know $g(0)\geq A_k$ and $g(c_k^{-1})\geq0$, it holds for $0\leq q\leq c_k^{-1}$ that
\[
g(q)
\geq
\left(1-q c_k\right)g(0)
+q c_k g(c_k^{-1})
\geq(1-c_kq)A_k.
\]
This is exactly \eqref{end:chord}.
\end{proof}

The preceding estimate gives a uniform gap strictly larger than $1$ throughout a large-separation parameter range:

\begin{corollary}[The large-separation range]
\label{end:large-k}
For $k\geq2/5$, $0\leq q\leq1$ and every $j\in\mathbb Z+\tfrac12$,
\begin{equation}\label{end:large-k-bound}
\nu_{k,j}(q)>1.
\end{equation}
\end{corollary}

\begin{proof}
Note that the right side of \eqref{end:chord} is a quadratic polynomial in $q$. Since $c_k$ decreases and $A_k$ increases with $k$, it suffices to bound this quadratic at $k=2/5$. We present this elementary (but technical estimate) in Appendix~\ref{app:large-scalar}, which completes the proof.
\end{proof}

\section{Preparation for the small-separation regime: the radial operator}
\label{sec:radial}

The majority of the remaining content of the paper is dedicated to extending the estimate of the gap function in Corollary \ref{end:large-k} to the small-$k$ regime. In particular, as remarked at the end of Section \ref{sec:axial}, the lower bound of the gap function is in fact controlled by the lowest angular-momentum sector of the axial operator, i.e., $H^{(3)}_{k,q,1/2}$. In this section, we present a decomposition of $H^{(3)}_{k,q,1/2}$ into its radial and angular parts, which is particularly convenient for the estimate of spectral gap. A similar decomposition is already seen in \eqref{ax:radial-forms}-\eqref{ax:angular-forms} for $q=0$, where its advantage is clear: the squared operator $(H^{(3)}_{k,0,1/2})^2$ becomes a sum of scalar operators acting separately in $\mu$ and $\nu$. Here we generalize the decomposition in \eqref{ax:radial-forms}-\eqref{ax:angular-forms} to $0\leq q\leq1$.

Throughout this section, we assume that $k>0$, $0\leq q\leq1$ and $j=1/2$. We decompose the operator $H^{(3)}_{k,q,1/2}$ as
\begin{equation}\label{rad:decomposition}
H^{(3)}_{k,q,1/2}=H_0+kW, \quad H_0:=R+A_0, \quad W:=B\sin\nu,
\end{equation}
where
\begin{align}
R&:=-i\zeta_3\sigma_1\partial_\mu +\zeta_3\sigma_3(k\tau_1\sinh\mu-q) +\tfrac12\zeta_1\tau_2\operatorname{csch}\mu,
\label{rad:operator}\\
A_0&:=-iS\partial_\nu+\tfrac12T\csc\nu, \quad
S:=\zeta_3\sigma_2, \quad T:=\zeta_1\tau_1, \quad B:=\zeta_3\sigma_3\tau_2.
\label{rad:angular-operator}
\end{align}
Note that the decomposition \eqref{rad:decomposition} follows our separation-of-variable guidelines: the operator $R$ acts on the radial variable $\mu$, while $A_0$ acts on the angular variable $\nu$; the remaining term $kW$ is a bounded angular perturbation, with controlled norm in the small-$k$ regime.

Here and thereafter, we suppress the $(k,q)$-dependence of $R$ and $H_0$ for ease of notation. In particular, $H_0=R+A_0$ retains the $q$-dependence through $R$. We follow the conventions of Pauli matrices at the beginning of Section \ref{sec:axial}, and write
\begin{equation} \label{eq:matrix_KN}
K:=\sigma_2\tau_1,
\quad
N:=\zeta_2\sigma_1\tau_2.
\end{equation}
The matrices $K$ and $N$ will be linked to the decomposition \eqref{rad:decomposition}-\eqref{rad:angular-operator} through the algebraic identity $B=TNK$. The following commutation/anti-commutation identities, understood on the test functions $C_{c}^{\infty}(\Omega;\mathbb{C}^{8})$, will be important for justifying the radial-angular separation property of \eqref{rad:decomposition}, which are verified by direct calculation
\begin{equation}\label{rad:matrix-relations}
\begin{gathered}
\{K,R\}=\{S,R\}=\{T,R\}=0,\\
[K,S]=[K,T]=[K,B]=[N,S]=[N,T]=0.
\end{gathered}
\end{equation}

The main result of this section focuses on the radial operator $R$. In particular, we characterize its kernel, the complement gap, and the action of $N$ on the kernel. These results will be applied later in Section \ref{sec:angular} to study the gap of $H^{(3)}_{k,q,1/2}$.

\begin{proposition}\label{rad:main}
The operator $R$ in \eqref{rad:operator} is essentially self-adjoint on $C_c^\infty((0,\infty);\mathbb C^8)$, and its closure has compact resolvent. The kernel $\ker R$ is two-dimensional and is contained in the sector $\ker(K-1)$. Let $P$ denote the orthogonal projection onto $\ker R$, and $Q:=I-P$. Then
\begin{equation}\label{rad:gap}
R^2\big|_Q\geq\mu_*(k,q),
\quad \text{with }\,
\mu_*(k,q):=q^2+2k\bigl(1+\sqrt{1-q^2}\bigr).
\end{equation}
Moreover, there exists a number $\rho=\rho(k,q)$ such that
\begin{equation}\label{rad:polarization-identities}
PNP=\rho P, \quad
PNQNP=(1-\rho^2)P, \quad
\frac{2k}{2k+q^2}\leq\rho\leq1.
\end{equation}
In particular, $\rho(k,0)=1$.
\end{proposition}

The three properties in Proposition~\ref{rad:main} have separate roles. The kernel of the radial operator $R$, together with the lowest positive angular eigenfunction, identifies the mode at the threshold $1$ of $H_0$. The bound on $R^2|_Q$ separates the remaining radial states from this threshold. Finally, $PNP=\rho P$ measures the part of the perturbation that stays in the radial kernel, while $PNQNP=(1-\rho^2)P$ controls its coupling to the radial complement.

The remainder of this section determines the kernel and spectral gap and proves the estimate for $\rho$. We retain the domain statements needed for these arguments, which are technical and do not contribute to the main understanding, and defer their proofs to Appendix~\ref{app:domains} as we did in Proposition \ref{ax:domain}.

\subsection{Self-adjointness and compactness}

\begin{lemma}\label{rad:domain}
The closure of $R$ in \eqref{rad:operator} is self-adjoint, with domain
\begin{equation}\label{rad:maximal-domain}
\Dom(R) = \bigl\{f\in L^2((0,\infty);\mathbb C^8): Rf\in L^2\text{ in distributions}\bigr\}.
\end{equation}
In particular, $\Dom(R)$, equipped with the graph norm, is compactly embedded in $L^2$. Consequently, the self-adjoint operator $R$ has compact resolvent.
\end{lemma}

Here and thereafter, with a little abuse of notation, we denote the self-adjoint closure of $R$ by the same symbol. The proof of Lemma~\ref{rad:domain} is given in Appendix~\ref{app:radial-domain}. For the spectral estimates below, we also record the square of $R$ on $C_c^\infty((0,\infty);\mathbb C^8)$:
\begin{align}
R^2=&-\partial_\mu^2 +\tfrac14\operatorname{csch}^2\mu -\tfrac12N\operatorname{csch}\mu\coth\mu +k^2\sinh^2\mu+q^2\nonumber\\
&-2kq\tau_1\sinh\mu-kK\cosh\mu-kNK.
\label{rad:square}
\end{align}

\subsection{The kernel and the positive radial gap} \label{sec:kernel_radial_gap}

\text{}

{\color{blue}Step 1. Reduction of the zero-mode equation.} To study the kernel of $R$, we first give an exact entry-wise characterization of the functions in $\ker R$. Suppose that $Rf=0$. Then multiplying by $i\zeta_3\sigma_1$ gives
\begin{equation}\label{rad:zero-equation}
f^{\prime}=
\left( -kK\sinh\mu+q\sigma_2 +\tfrac12N\operatorname{csch}\mu \right)f.
\end{equation}
Note that all coefficients in \eqref{rad:zero-equation} commute with $K$ and $\zeta_2$. We may therefore work in a joint eigenspace
with eigenvalues $\varepsilon,c\in\{-1,1\}$ of $K,\zeta_2$, respectively, with each of the four subspaces being two-dimensional. Next, we write
\begin{equation} \label{eq:K=1_subspace}
f=e^{-\varepsilon k\cosh\mu}F.
\end{equation}
Since the matrices $\sigma_2$ and $N$ are anticommuting Hermitian involutions on each subspace indexed by $(\varepsilon,c)$, we may choose a specific representation
\begin{equation} \label{eq:K=1_subspace_pauli}
\sigma_2=
\begin{pmatrix}0&1\\1&0\end{pmatrix}:=\eta_1,
\quad
N=
\begin{pmatrix}1&0\\0&-1\end{pmatrix}:=\eta_3.
\end{equation}
Then $F=(u,v)^{\mathsf T}$ satisfies
\begin{equation}\label{rad:ode}
u^{\prime}=qv+\frac{u}{2\sinh\mu},
\quad
v^{\prime}=qu-\frac{v}{2\sinh\mu}.
\end{equation}

{\color{blue}Step 2. Unique regular solution of \eqref{rad:ode}.}

{\color{blue}Step 2.1.} We now construct a regular solution of \eqref{rad:ode} with normalization
\begin{equation}\label{rad:regular-solution}
u(\mu)\sim\sqrt\mu, \quad
v(\mu)= \tfrac q2\mu^{3/2} +O(\mu^{7/2}) \quad(\mu\to 0^{+}),
\end{equation}
and the boundary condition $v(0)=0$. In particular, the system \eqref{rad:ode} is equivalent to the Volterra equations
\begin{equation} \label{eq:volterra}
V(\mu)=q\int_0^\mu a(s)^2U(s)\,ds, \quad
U(\mu)=\sqrt2+q\int_0^\mu a(t)^{-2}V(t)\,dt,
\end{equation}
with the substitution
\begin{equation*}
a(\mu):=\sqrt{\tanh(\mu/2)},\quad
U(\mu):=\frac{u(\mu)}{a(\mu)},\quad
V(\mu):=a(\mu)v(\mu),
\end{equation*}
and noting that $a'/a=1/(2\sinh\mu)$. Then a fixed-point iteration gives a unique non-trivial solution to \eqref{eq:volterra} for small $\mu$, and hence \eqref{rad:ode}, which extends to all $\mu>0$ by the well-posedness of the ODE system \eqref{rad:ode}. In particular, since the kernels in \eqref{eq:volterra} are nonnegative, we know that $u>0$ and $v\geq 0$ for all $\mu>0$. Moreover, we claim that
\begin{equation} \label{eq:zero_mode_dominance}
u>v\geq 0,\quad \text{for all $\mu>0$}.
\end{equation}
Indeed, \eqref{rad:ode} gives
\[
\frac{d}{d\mu}\bigl[e^{q\mu}(u-v)\bigr]
=\frac{e^{q\mu}(u+v)}{2\sinh\mu}>0.
\]
Hence, the claim is justified by recalling that $u-v>0$ for small $\mu>0$.

{\color{blue}Step 2.2.} Next, we prove that the regular solution constructed in the previous step is the only local $L^2$ solution of \eqref{rad:ode}. Suppose $\tilde{F}$ is another solution to \eqref{rad:ode} such that
\[
\widetilde F\in L^2((0,\delta);\mathbb C^2)
\quad\text{for some }\delta>0.
\]
First of all, since the coefficient matrix in \eqref{rad:ode} has trace zero, we know the Wronskian associated with $(F,\tilde{F})$ is constant. On the other hand, as in Step 1 of Appendix~\ref{app:radial-domain}, the negative component $\tilde{F}_{-}:=(I-N)\tilde{F}/2$ has the asymptotics $o(\sqrt\mu)$ near zero, while the positive component $\tilde{F}_{+}:=(I+N)\tilde{F}/2$ is $o(\mu^{-1/2})$ along a sequence approaching zero. Hence, using \eqref{rad:regular-solution}, the Wronskian associated with $(F,\tilde{F})$ tends to zero along that sequence, which proves the linear dependence $\tilde{F}\propto F$.

{\color{blue}Step 2.3.} We next examine the asymptotic behavior of the solution $F$ at infinity. For $\mu\geq1$, the coefficient matrix in \eqref{rad:ode} has norm at most $q+\tfrac12\operatorname{csch}\mu$. Since $\operatorname{csch}\mu$ is integrable at infinity, the Gronwall estimates give
\[
c_F e^{-q\mu}\leq|F(\mu)|\leq C_F e^{q\mu}
\]
with $c_F,C_F>0$. It follows that $e^{-k\cosh\mu}F$ is square integrable at infinity, whereas $e^{k\cosh\mu}F$ is not. This implies that one must take $\varepsilon=1$ in \eqref{eq:K=1_subspace} to guarantee an integrable $f$. In other words, a zero mode $f$ must lie in the $K=1$ subspace. Finally, the additional two degrees of freedom given by the index $c\in\{\pm 1\}$ leads to two independent solutions. These two solutions belong to $\operatorname{Dom}(R)$ by \eqref{rad:maximal-domain}, since they are in $L^2$ and satisfy $Rf=0$ distributionally, by which we conclude that
\[
\dim\ker R=2,
\qquad
\ker R\subset\{K=1\}.
\]

{\color{blue}Step 3. The radial complement gap.}

In this step, we prove the gap estimate \eqref{rad:gap}. First, by the anti-commutation relation $KR=-RK$, we may write $R$ into the following block form adapted to the $K=1$ and $K=-1$ sectors:
\[
R=
\begin{pmatrix}
0&\mathcal L^*\\
\mathcal L&0
\end{pmatrix}, \quad
R^2=
\begin{pmatrix}
\mathcal L^*\mathcal L&0\\
0&\mathcal L\mathcal L^*
\end{pmatrix},
\]
where $\mathcal L$ is closed and densely defined. The result from Step 2 indicates that
\[
\dim\ker\mathcal L=2, \quad \ker\mathcal L^*=\{0\}.
\]
By the polar decomposition of $\mathcal L$, $\mathcal L^*\mathcal L$ on $(\ker\mathcal L)^\perp$ is unitarily equivalent to $\mathcal L\mathcal L^*$. It therefore suffices to prove the desired lower bound \eqref{rad:gap} on the $K=-1$ subspace. Remarkably, $\tau_1=-\sigma_2$ on this subspace. Hence, applying the representation \eqref{eq:K=1_subspace_pauli}, the squared operator $R^2$ in \eqref{rad:square} becomes
\begin{equation} \label{eq:radial_gap_proof_1}
\begin{aligned}
R^2=\, &q^2+2kq\eta_1\sinh\mu+k\cosh\mu\\
&+\left[
-\partial_\mu^2+\tfrac14\operatorname{csch}^2\mu
-\tfrac12\eta_3\operatorname{csch}\mu\coth\mu
+k^2\sinh^2\mu+k\eta_3
\right].
\end{aligned}
\end{equation}
{\color{blue}Step 3.1.} We claim that
\begin{equation}
\label{eq:radial_gap_proof_2}
-\partial_\mu^2+\tfrac14\operatorname{csch}^2\mu -\tfrac12\eta_3\operatorname{csch}\mu\coth\mu
+k^2\sinh^2\mu+k\eta_3
\geq k\cosh\mu+2k .
\end{equation}
To prove this, define
\[
h_\pm:=
-\partial_\mu^2+k^2\sinh^2\mu
+\tfrac14\operatorname{csch}^2\mu
\mp\tfrac12\operatorname{csch}\mu\coth\mu.
\]
Then the difference of two sides of \eqref{eq:radial_gap_proof_2} becomes $h_+-k\cosh\mu-k$ when $\eta_3=1$, and $h_- -k\cosh\mu-3k$ when $\eta_3=-1$. In the $\eta_3=1$ component, \eqref{eq:radial_gap_proof_2} is proved directly by factorization
\[
\begin{aligned}
h_+-k\cosh\mu-k
&=
\left(
\partial_\mu+k\sinh\mu-\tfrac12\operatorname{csch}\mu
\right)^*
\left(
\partial_\mu+k\sinh\mu-\tfrac12\operatorname{csch}\mu
\right)\\[-2pt]
&\geq0.
\end{aligned}
\]
In the $\eta_3=-1$ component, choose $f(\mu):=e^{-k\cosh\mu}\tanh(\mu/2)^{3/2}$ and directly calculate that
\[
\frac{(h_- -k\cosh\mu-3k)f}{f}
=\frac2{\cosh\mu+1}>0.
\]
Thus the ground state estimate \eqref{eq:ground_state_estimate} yields $h_-\geq k\cosh\mu+3k$. This completes the proof of \eqref{eq:radial_gap_proof_2}.

{\color{blue}Step 3.2.} By \eqref{eq:radial_gap_proof_2}, we know from \eqref{eq:radial_gap_proof_1} that
\[
R^2\big|_{K=-1} \geq q^2+2k+2k(\cosh\mu-q\sinh\mu),
\]
where we have applied $\eta_1\geq-I$ and $q\geq0$. In particular, for $0\leq q\leq1$,
\[
\cosh\mu-q\sinh\mu\geq\sqrt{1-q^2}.
\]
Hence,
\[
R^2\big|_{K=-1} \geq q^2+2k\bigl(1+\sqrt{1-q^2}\bigr) =\mu_*(k,q).
\]
The inequalities are first obtained on the test functions and extend to the squared quadratic form of $R$ by graph-core approximation, provided by Lemma \ref{rad:domain}. This proves \eqref{rad:gap}.

\subsection{A lower bound for the radial polarization}

Finally, we prove the identities \eqref{rad:polarization-identities}. Intuitively speaking, the factor $\rho$ arises as the expectation of matrix $N$ in the zero modes. Recall that the two normalized zero modes constructed in Step 2 of Section \ref{sec:kernel_radial_gap} are copies of $e^{-k\cosh\mu}(u,v)^{\mathsf T}$, with $u$ and $v$ satisfying \eqref{rad:ode}, in the two $\zeta_2$-sectors. Since $N$ commutes with $\zeta_2$, the off-diagonal elements of its matrix in the basis of zero modes vanish. On the other hand, the expectations of $N$ in the two zero modes are identical, and both equal
\begin{equation}\label{rad:rho}
\rho(k,q):=
\frac{\displaystyle
\int_0^\infty e^{-2k\cosh\mu}(u^2-v^2)\,d\mu}
{\displaystyle
\int_0^\infty e^{-2k\cosh\mu}(u^2+v^2)\,d\mu}.
\end{equation}
With this notation, the identity $PNP=\rho P$ follows immediately. Moreover, the estimate \eqref{eq:zero_mode_dominance} implies $0<\rho\leq1$. Since $N^2=I$ and $Q=I-P$, one checks that
\[
PNQNP=PN(I-P)NP=P-(PNP)^2=(1-\rho^2)P.
\]
Thus, it remains to prove the quantitative lower bound of $\rho$ claimed in \eqref{rad:polarization-identities}:

\begin{lemma}\label{rad:rho-bound}
For every $k>0$ and $q\in[0,1]$,
\begin{equation}\label{rad:rho-lower}
\rho(k,q)\geq\rho_0(k,q):=\frac{2k}{2k+q^2}.
\end{equation}
Equality holds at $q=0$ and $q=1$.
\end{lemma}

\begin{proof}
The equality of \eqref{rad:rho-lower} is clear at $q=0$ as $v=0$ therein. In the sequel, we assume $0<q\leq1$. The proof is based on the characterization of the zero modes in Section \ref{sec:kernel_radial_gap}. Specifically, we will first give an estimate of the ratio $v/u$, then apply it to estimate the ratio $(u^2-v^2)/(u^2+v^2)$, by which we conclude the proof using a probabilistic estimate of expectations.

{\color{blue}Step 1. A comparison for the ratio $v/u$.}

Thanks to the estimate $u>v\geq0$ in \eqref{eq:zero_mode_dominance}, the ratio $r:=v/u$ is well-defined and satisfies $0\leq r<1$. By \eqref{rad:ode},
\[
r^{\prime}=q(1-r^2)-\operatorname{csch}\mu\,r,
\quad
r(\mu)=\tfrac q2\mu+O(\mu^3)
\quad(\mu\to 0^{+}).
\]
In this step, we compare $r$ with an explicit super-solution. Write
\[
\overline r:=\frac{qz}{y},\quad \text{with }\,
z:=\tanh(\mu/2),\, y:=\sqrt{1-(1-q^2)z^2}.
\]
Direct differentiation gives
\begin{equation}\label{rad:riccati-residual}
\overline r^{\prime}-q(1-\overline r^2) +\operatorname{csch}\mu\,\overline r
= \frac{q(1-z^2)(1-y)^2}{2y^3}\geq0.
\end{equation}
We claim that
\begin{equation} \label{eq:supersolution}
r\leq\overline r.
\end{equation}
Indeed, letting $\delta:=\overline r-r$, we know
\[
\delta^{\prime} +\bigl[q(\overline r+r)+\operatorname{csch}\mu\bigr]\delta
=\frac{q(1-z^2)(1-y)^2}{2y^3}.
\]
A positive integrating factor for this equation is comparable to $\mu$ near zero. Since $\delta=O(\mu)$, the corresponding endpoint term at $\mu=0$ vanishes. Integration
from zero therefore gives $\delta\geq0$ and proves \eqref{eq:supersolution}.

{\color{blue}Step 2. Estimates of $(u^2-v^2)/(u^2+v^2)$.}

Let $\mathcal S:=u^2+v^2$ and $\mathcal D:=u^2-v^2$, which are all positive for $\mu> 0$ and vanish at $\mu=0$. Since 1) $1-\overline r^2=\frac{1-z^2}{y^2}>0$ and 2) $r\mapsto(1+r^2)/(1-r^2)$ is increasing on
$[0,1)$, \eqref{eq:supersolution} leads to
\begin{equation}\label{rad:pointwise-polarization}
\frac{\mathcal S}{\mathcal D}
=\frac{1+r^2}{1-r^2}
\leq
\frac{1+\overline r^2}{1-\overline r^2}
=1+q^2(\cosh\mu-1).
\end{equation}
On the other hand, using \eqref{rad:ode}, one can calculate that
\[
\mathcal D'=\frac{\mathcal S}{\sinh\mu}.
\]
Consequently,
\[
(\log\mathcal D)'
\leq
\frac{1+q^2(\cosh\mu-1)}{\sinh\mu}
=
\frac{d}{d\mu}
\log\bigl[\sinh\mu(\cosh\mu+1)^{q^2-1}\bigr],
\]
which implies
\begin{equation}\label{rad:pointwise-polarization_alternative}
\frac{\mathcal D(\mu)}
{\sinh\mu(\cosh\mu+1)^{q^2-1}}
\quad \text{ is non-increasing.}
\end{equation}

{\color{blue}Step 3. Integration of the pointwise estimate.}

We use the following elementary comparison of probability densities to conclude the proof of \eqref{rad:rho-lower}: if $p_1,p_2$ are positive normalized densities and $p_1/p_2$ is non-increasing, then
\[
\int F\cdot p_1\leq\int F\cdot p_2
\]
for every increasing function $F$ with finite expectations. Indeed, for $\ell:=p_1/p_2$, it holds that
\[
\int Fp_1-\int Fp_2
=\operatorname{Cov}_{p_2}(F,\ell)\leq0,
\]
since $F$ is increasing and $\ell$ is non-increasing. Applying this comparison to $F=\cosh\mu-1$ with normalized densities
\[
p_1\propto e^{-2k\cosh\mu}\mathcal D(\mu)
\quad\text{and}\quad
p_2\propto e^{-2k\cosh\mu}\sinh\mu(\cosh\mu+1)^{q^2-1},
\]
whose ratio is non-increasing by \eqref{rad:pointwise-polarization_alternative}, we know
\begin{equation*}
\mathbb E_{p_1}(\cosh\mu-1)\leq \mathbb E_{p_2}(\cosh\mu-1)\leq \frac{1}{2k},
\end{equation*}
where the second inequality follows by direct calculation using the change of variables $Y=\cosh\mu-1$. With this estimate and taking the expectation in \eqref{rad:pointwise-polarization} with respect to the normalized density $p_1$, we obtain
\[
\rho^{-1}=
\mathbb{E}_{p_1} \left(\frac{\mathcal S}{\mathcal D}\right)
\leq 1+q^2\mathbb{E}_{p_1}(\cosh\mu-1) \leq1+\frac{q^2}{2k}.
\]
This proves the inequality part of \eqref{rad:rho-lower}.

{\color{blue}Step 4.} It only remains to prove the equality of \eqref{rad:rho-lower} at $q=1$. In fact, when $q=1$, we have $y=1$, so the residual in \eqref{rad:riccati-residual} vanishes. Then the integrating-factor argument in Step 1 gives $r=\overline r$, whereas both \eqref{rad:pointwise-polarization} and \eqref{rad:pointwise-polarization_alternative} become equalities. Moreover, using the fact $q^2=1$, one can also calculate that $\mathbb{E}_{p_2}(\cosh\mu-1)= \frac{1}{2k}$; therefore, all inequalities in Step 3 are equalities for $q=1$, which concludes the proof.
\end{proof}

\subsection{The separated operator and its domain}

We finish this section by specifying the domain of the separated operator in \eqref{rad:decomposition}, especially the angular part $A_0$ in \eqref{rad:angular-operator}. This justifies the operator decompositions used in the next section.

\begin{lemma}\label{rad:sum-domain}
The angular expression \eqref{rad:angular-operator} is essentially self-adjoint on $C_c^\infty((0,\pi);\mathbb C^8)$. Its closure has the maximal distributional domain. The operator $H_0=R+A_0$ is self-adjoint on $\Dom(R)\cap\Dom(A_0)$ and satisfies
\begin{equation}\label{rad:graph-identity}
\|H_0f\|^2=\|Rf\|^2+\|A_0f\|^2.
\end{equation}
Here, $R$ and $A_0$ are understood as operators in their respective variables on $L^2(\Omega;\mathbb C^8)$. Moreover, $C_c^\infty(\Omega;\mathbb C^8)$ forms a graph core for $H_0$. The spectral projections of $R^2$ in Proposition \ref{rad:main} commute with $A_0$, and hence
\begin{equation} \label{eq:strong_commute}
H_0^2=R^2+A_0^2
\end{equation}
holds in the joint spectral sense.
\end{lemma}

The proof is given in Appendix~\ref{app:separated-domain}. Since $kW$ is bounded, the same domain and graph core apply to $H_0+kW$. Thus, the decomposition \eqref{rad:decomposition} holds as an identity of self-adjoint operators on $\Dom(H_0)$.

\section{An angular Schur complement: conditional small-separation gap estimate}
\label{sec:angular}

In this section, we apply the characterization of the radial operator $R$ obtained in the last section, together with the spectrum of the angular operator that can be calculated explicitly (see Section \ref{sec:angular_spectrum}), to estimate the gap of $H^{(3)}_{k,q,1/2}$ in the small-$k$ regime. Specifically, this is achieved by projecting $H^{(3)}_{k,q,1/2}$ onto the lowest positive angular mode in $\ker R$ and estimating the resulting scalar Schur complement. In particular, as we will show, an exact angular selection rule reduces the complement vector to three mutually orthogonal terms, which leads to the estimation of gap of $H^{(3)}_{k,q,1/2}$ conditioned on an explicit function; see Sections \ref{sec:complement_resolvent} and \ref{sec:proof_angular_criteria} for details.


Throughout this section, we retain the notation of Section \ref{sec:radial} and assume that $0<k\leq 2/5$ and $0\leq q\leq1$. We define
\begin{equation}\label{ang:parameters}
\begin{gathered}
d_0(k,q)=\sqrt{1+\mu_*}-1,\quad
\rho_0(k,q)=\frac{2k}{2k+q^2},
\end{gathered}
\end{equation}
where $\mu_*(k,q)$ is introduced in \eqref{rad:gap}. For $0\leq r\leq1$ and $h>0$, we also write
\begin{align}
M(r,h)
&:=\frac{r^2}{18} +\frac49(1-r^2)(h^{-1}+4h^{-2}) +\frac{2(1-r^2)}{9(h+4)},
\label{ang:M}\\
F(k,q)
&:=\frac4{3(2k+q^2)}
-\frac{8(1-\rho_0^2)}{9\mu_*}
-\frac{k}{1-k/d_0}M(\rho_0,\mu_*).
\label{ang:F}
\end{align}
The main result of this section gives a sufficient condition for the desired spectral gap conditioned on the positivity of the explicit function $F$. In Section~\ref{sec:bridge}, we combine this criterion with the endpoint estimate to complete the parameter coverage.

\begin{proposition} \label{ang:criterion}
For $0<k\leq2/5$ and $0\leq q\leq1$, the denominators in \eqref{ang:F} are positive. If $F(k,q)>0$, then
\begin{equation}\label{ang:gap}
\operatorname{spec}(H^{(3)}_{k,q,1/2})\cap[-1,1]
=\varnothing.
\end{equation}
\end{proposition}

\subsection{The angular spectrum and a selection rule} \label{sec:angular_spectrum}

By \eqref{rad:matrix-relations} and Proposition \ref{rad:main}, the two-dimensional kernel of $R$, i.e., $\text{Ran}(P)$, is invariant under the matrices $S$ and $T$, with their restrictions on the kernel satisfying
\[
S^*=S,\quad T^*=T,\quad S^2=T^2=I,\quad ST=-TS.
\]
We may therefore choose an orthonormal basis of $\ker R$ in which $S,T$ are represented by the Pauli matrices $S=\sigma_2$ and $T=\sigma_1$. Under this representation, the restriction of the angular operator $A_0$ becomes
\begin{equation}\label{ang:A}
\mathcal A=
\begin{pmatrix}
0&-\partial_\nu+(2\sin\nu)^{-1}\\
\partial_\nu+(2\sin\nu)^{-1}&0
\end{pmatrix},
\quad 0<\nu<\pi.
\end{equation}
We also write
\[
V:=\sigma_1\sin\nu,
\]
which represents the restriction of $T\sin\nu$. In fact, the operator $T\sin\nu$ arises naturally when we consider the compression of the perturbation operator $W$ to the kernel:
\[
 PWP=PTNP\sin\nu=\rho\,T\sin\nu\,P.
\]
Indeed, the definitions \eqref{rad:angular-operator} and \eqref{eq:matrix_KN} give $B=TNK$, while $\ker R\subset\ker(K-I)$ implies $KP=P$. Hence $BP=TNP$, and the displayed identity follows from $PT=TP$ and $PNP=\rho P$.

\begin{lemma}\label{ang:spectrum}
The spectrum of $\mathcal A$ consists of the simple eigenvalues $\pm n$, with $n\in\mathbb{N}$. Consequently, $A_0^2\geq1$ on the full spinor space $L^2(\Omega;\mathbb{C}^{8})$. The eigenmode of $\mathcal A$ associated with the eigenvalue $1$ is
\begin{equation}\label{ang:ground}
\psi_1(\nu)=\frac12
\begin{pmatrix}
\sin\nu\sqrt{\tan(\nu/2)}\\
\sin\nu\sqrt{\cot(\nu/2)}
\end{pmatrix}.
\end{equation}
Moreover, there exists a normalized eigenvector $\psi_{-2}$ with
eigenvalue $-2$ such that
\begin{equation}\label{ang:selection}
V\psi_1=\frac23\psi_1+\frac{\sqrt2}{3}\psi_{-2}.
\end{equation}
\end{lemma}

\begin{proof}

We will first study the spectrum of $\mathcal A^2$, which is explicitly computable using Jacobi polynomials. Then the spectrum of $\mathcal{A}$ as well as the claim on its eigenmodes follows by exploiting the symmetry of \eqref{ang:A}.

{\color{blue}Step 1. Spectrum of $\mathcal A^2$.} Taking the square of \eqref{ang:A} gives
\[
\mathcal A^2=
\begin{pmatrix}
L_{-1}&0\\
0&L_1
\end{pmatrix},
\quad \text{with }\,
L_\varepsilon:=
-\partial_\nu^2+\tfrac14\csc^2\nu
-\tfrac{\varepsilon}{2}\csc\nu\cot\nu
\quad (\varepsilon\in\{-1,1\}).
\]
Let $p_{n,1}:=P_n^{(0,1)}$ and $p_{n,-1}:=P_n^{(1,0)}$ be the Jacobi polynomials with the indicated parameters for $n\geq 0$, which satisfy
\[
-(1-x^2)p_{n,\varepsilon}''
+(3x-\varepsilon)p_{n,\varepsilon}'
=n(n+2)p_{n,\varepsilon} \quad (\varepsilon\in\{-1,1\}).
\]
Consequently, the functions $f_{n,\varepsilon}(\nu)
:=f_{0,\varepsilon}(\nu)p_{n,\varepsilon}(\cos\nu)$ with $f_{0,\varepsilon}(\nu):=\sin\nu\,\tan(\nu/2)^{-\varepsilon/2}$ satisfy
\[
L_\varepsilon f_{n,\varepsilon}
=(n+1)^2f_{n,\varepsilon}.
\]
To complete the proof of the claim on $\text{spec}(\mathcal{A})$, it remains to show that 1) 
\begin{equation} \label{eq:mathcalA_eigenfunction_domain}
\Phi_{n,\varepsilon}\in\operatorname{Dom}(\mathcal A^2),\quad
\text{with }\,
\Phi_{n,-1}:=
\begin{pmatrix}f_{n,-1}\\0\end{pmatrix},\,
\Phi_{n,1}:=
\begin{pmatrix}0\\f_{n,1}\end{pmatrix},
\end{equation}
and 2) $\text{spec}(\mathcal A^2)$ consists of discrete eigenvalues. For the first point, note that the regularity of Jacobi polynomials directly gives
\[
\Phi_{n,\varepsilon}\in L^2,\quad
\mathcal A\Phi_{n,\varepsilon}\in L^2,\quad
\mathcal A(\mathcal A\Phi_{n,\varepsilon})
=(n+1)^2\Phi_{n,\varepsilon}\in L^2
\]
in the distributional sense. The maximal-domain characterization in Lemma \ref{rad:sum-domain} therefore implies \eqref{eq:mathcalA_eigenfunction_domain}. For the second point, note that the Jacobi polynomials $p_{n,\varepsilon}$ form a complete orthogonal basis in $L^2((-1,1),(1+\varepsilon x)\,dx)$ for each $\varepsilon\in\{-1,1\}$; hence, by the unitarity
\[
\int_0^\pi \left|f_{0,\varepsilon}(\nu)g(\cos\nu)\right|^2\,d\nu
=\int_{-1}^{1}|g(x)|^2(1+\varepsilon x)\,dx,
\]
we know that these spinors form a complete system of eigenvectors of $\mathcal A^2$ and thus exhaust its spectrum. Moreover, the preceding argument also implies that each eigenvalue of $\mathcal A^2$ is two-fold.

Since 1) $\mathcal A^2\geq 1$ as proved above, and 2) the Hermitian involutions \(S,T\), satisfying \(ST=-TS\), decompose \(\mathbb C^8\) into four equivalent two-dimensional representations, we conclude that the full angular operator $A_0$ is unitarily equivalent to four copies of \(\mathcal A\), and therefore, \(A_0^2\ge1\).

{\color{blue}Step 2. The angular spectrum.} Note that \eqref{ang:A} exhibits the particle-hole-like symmetry:
\[
\mathcal A\sigma_3=-\sigma_3\mathcal A,
\]
by which we know that each two-fold eigenvalue $(n+1)^2$ of $\mathcal{A}^2$ splits into the two simple eigenvalues $\pm(n+1)$ of $\mathcal A$.

{\color{blue}Step 3. Proof of \eqref{ang:selection}.} A direct calculation using \eqref{ang:ground} gives
\[
\mathcal A\psi_1=\psi_1,
\quad
|\psi_1(\nu)|^2=\tfrac12\sin\nu,
\quad
\|\psi_1\|=1,
\]
and
\[
\langle\psi_1,V\psi_1\rangle
=\|V\psi_1\|^2=\frac23,
\quad
\mathcal AV\psi_1=2\psi_1-2V\psi_1.
\]
Hence, by defining
\[
\psi_{-2}:=\frac{3V\psi_1-2\psi_1}{\sqrt2},
\]
we can directly check that $\psi_{-2}$ is an eigenmode of $\mathcal{A}$ associated with the eigenvalue $-2$, which completes the proof of \eqref{ang:selection}.
\end{proof}

\subsection{The complement resolvent} \label{sec:complement_resolvent}

Denote $\iota:\mathbb C^2\to\ker R$ by the isometric identification with the two-dimensional kernel of $R$, which has been applied implicitly in Section \ref{sec:angular_spectrum} and chosen to adapt the representation
\[
\iota^{-1}S\iota=\sigma_2,\quad
\iota^{-1}T\iota=\sigma_1.
\]
We use the same notation for its extension to angular functions and define
\[
\Psi_1:=\iota\psi_1,\quad
\Psi_{-2}:=\iota\psi_{-2},\quad
\Pi:=|\Psi_1\rangle\langle\Psi_1|.
\]
In particular,
\[
H_0\Psi_1=\Psi_1,\quad H_0\Psi_{-2}=-2\Psi_{-2}.
\]
In this section, we establish the resolvent estimate of $H_{k,q,1/2}^{(3)}$ adapted to the Schur complement associated with $\text{Ran}(\Pi)$ and its orthogonal complement. To this end, we first calculate the expectation of the perturbation operator $W$ (recall the decomposition \eqref{rad:decomposition}) in the eigenstate $\Psi_1$ of $H_0$. Let
\[
\rho:=\rho(k,q),\quad v:=1-\rho^2,
\]
with $\rho(k,q)$ introduced in Proposition \ref{rad:main}. We claim
\begin{equation} \label{ang:diagonal}
\langle\Psi_1,W\Psi_1\rangle=\frac23\rho.
\end{equation}
In fact, if we define
\begin{equation}\label{ang:C}
C:=QN\iota:
\mathbb C^2\longrightarrow
QL^2((0,\infty);\mathbb C^8),
\end{equation}
where $Q=I-P$ is also defined in Proposition \ref{rad:main}, and the matrix $N$ is defined in \eqref{eq:matrix_KN}, then one can check that
\begin{equation} \label{eq:perturbation_restriction}
W\iota=(\rho\iota+C)V.
\end{equation}
Taking the expectation in $\psi_1$ on both sides of \eqref{eq:perturbation_restriction} leads to \eqref{ang:diagonal}. The identity \eqref{eq:perturbation_restriction} is justified by noting the following identities
\[
N\iota=\rho\iota+C,\quad
C^*C=vI,\quad
KC=C,
\]
and
\[
SC=C\sigma_2,\quad TC=C\sigma_1,
\]
which are derived from Proposition \ref{rad:main}. Consequently, $A_0C=C\mathcal A$. Moreover, since $B=TNK$ and $K\iota=\iota$, we arrive at \eqref{eq:perturbation_restriction}.

On the other hand, we note that the off-diagonal component of the perturbation $W$ corresponding to the decomposition induced by $\Pi$ is given by
\begin{equation}\label{ang:Xi}
\Xi:=(I-\Pi)W\Psi_1
=\frac{\sqrt2}{3}\rho\Psi_{-2}
+\frac23C\psi_1
+\frac{\sqrt2}{3}C\psi_{-2},
\end{equation}
where we have applied \eqref{ang:selection} and \eqref{ang:diagonal} to derive the second equality. Importantly, the three vectors on the right side are orthogonal. This is proved by noting that 1) $R\Psi_{-2}=0$, while $C\psi_1,C\psi_{-2}\in \ker(R)^{\perp}$ by definitions, and 2) $C\psi_1$ and $C\psi_{-2}$ are mutually orthogonal as
\[
A_0C\psi_1=C\psi_1,\quad
A_0C\psi_{-2}=-2C\psi_{-2},
\]
and
\[
\|C\psi_1\|^2=\|C\psi_{-2}\|^2=v.
\]
Note that we do not assume $v\neq 0$ (indeed, $v=0$ holds for $q=0$) as in that case the claimed orthogonality still holds, with $C\psi_1=C\psi_{-2}=0$.

With these preparations, we are now ready to estimate the resolvent on the complement of $\Psi_1$. Fix $\lambda\in [0,1]$, let $h:=\mu_*(k,q)$ and
\begin{equation}\label{ang:delta}
\delta:=\min\{1,\sqrt{1+h}-1\}.
\end{equation}
In the sequel, the orthogonal complement of $\text{Ran}(\Pi)$ is denoted by
\[
\mathscr H_c:=(I-\Pi)L^2(\Omega;\mathbb C^8),
\]
and let $H_{0,c}$ be the restriction of $H_0$ to $\mathscr H_c$. Since $\Pi$ reduces $H_0$, this restriction is self-adjoint on $\operatorname{Dom}(H_0)\cap\mathscr H_c$. We claim that
\begin{equation}\label{ang:distance}
\operatorname{dist}
\bigl(\lambda,\operatorname{spec}H_{0,c}\bigr)
\geq\delta.
\end{equation}
Indeed, on $\ker(R)$, the remaining eigenvalues of $A_0$ have a distance at least one from $[0,1]$ by Lemma \ref{ang:spectrum}. On the other hand, Lemma \ref{rad:sum-domain} gives $H_0^2=R^2+A_0^2\geq h+1$ on the complement $\ker(R)^{\perp}$. This concludes the proof of \eqref{ang:distance}.

Now we present the resolvent estimate when including the perturbation $W$. Assume for the moment that $k<\delta$, and define on $\mathscr H_c$
\[
G_0:=(H_{0,c}-\lambda)^{-1},\quad
W_c:=(I-\Pi)W(I-\Pi),\quad
G_k:=(H_{0,c}+kW_c-\lambda)^{-1}.
\]
Both resolvents exist by recalling the bounds
\[
\|G_0\|\leq\delta^{-1},\quad
\|W_c\|\leq1,\quad
\|kG_0W_c\|\leq k/\delta<1,
\]
which follow from the estimate presented above. The following estimates are key for the Schur complement:

\begin{lemma}\label{ang:resolvent}
For $0\leq\lambda\leq1$ and $k<\delta$,
\begin{align}
\langle\Xi,G_0\Xi\rangle
&\leq\frac{8v}{9h},
\label{ang:first-loss}\\
\|G_0\Xi\|^2
&\leq M(\rho,h),
\label{ang:squared-loss}\\
\bigl|\langle\Xi,(G_k-G_0)\Xi\rangle\bigr|
&\leq\frac{k}{1-k/\delta}M(\rho,h),
\label{ang:perturbation-loss}
\end{align}
where the function $M$ is introduced in \eqref{ang:M}.
\end{lemma}

\begin{proof}

{\color{blue}Step 1. Diagonal resolvent on the $K=1$ subspace.}

We first present a formula of the diagonal component of the (unperturbed) resolvent $G_0$ on the $K=1$ subspace. Since $H_0^2-\lambda^2=R^2+A_0^2-\lambda^2\geq h>0$ on the radial complement $\text{Ran}(Q)$ by \eqref{rad:gap} and Lemma \ref{ang:spectrum}, we apply the identity $a^2-b^2=(a-b)(a+b)$ and write
\[
G_0=(R+A_0+\lambda)(R^2+A_0^2-\lambda^2)^{-1}\quad \text{on $\text{Ran}(Q)$.}
\]
Let $P_{\pm}:=(I\pm K)/2$. The relations
\[
RP_+=P_-R,\quad
[A_0,P_+]=[R^2+A_0^2-\lambda^2,P_+]=0
\]
show that the $K=1$ diagonal block of $G_0$ is
\begin{equation}\label{ang:diagonal-resolvent}
P_+G_0P_+
=(A_0+\lambda)
(R^2+A_0^2-\lambda^2)^{-1}P_+.
\end{equation}
Similarly,
\[
P_+G_0^2P_+
=
\frac{R^2+(A_0+\lambda)^2}
{(R^2+A_0^2-\lambda^2)^2}P_+.
\]
Here, the expressions are defined by the joint spectral calculus of $R^2$ and $A_0$ on the radial complement, as justified by Lemma \ref{rad:sum-domain}.

{\color{blue}Step 2.} We next prove \eqref{ang:first-loss}. On a common eigenvector of $R^2$ with eigenvalue $r\geq h$ and of $A_0$ with eigenvalue $a$, the right side of \eqref{ang:diagonal-resolvent} becomes a multiplier equal to
\[
\frac{a+\lambda}{r+a^2-\lambda^2}.
\]
Then $\langle\Xi,G_0\Xi\rangle$ equals the sum of multipliers associated with the three vectors in \eqref{ang:Xi}, due to their orthogonality. In particular, the two terms with angular eigenvalue $-2$ in \eqref{ang:Xi} give non-positive contributions. For the
term $C\psi_1$, we have
\[
\frac{1+\lambda}{r+1-\lambda^2}
\leq\frac2r\leq\frac2h.
\]
Consequently,
\[
\langle\Xi,G_0\Xi\rangle
\leq\frac49\frac2h\|C\psi_1\|^2
=\frac{8v}{9h},
\]
which proves \eqref{ang:first-loss}.

{\color{blue}Step 3.} We next prove \eqref{ang:squared-loss}. Similarly to the argument in the previous step, we point out that the corresponding multiplier for the $K=1$ diagonal block of $G_0^2$ is
\[
\frac{r+(a+\lambda)^2}{(r+a^2-\lambda^2)^2}.
\]
For $0\leq\lambda\leq1$ and $r>0$,
\[
\frac{r+(1+\lambda)^2}{(r+1-\lambda^2)^2} \leq\frac{r+4}{r^2} =\frac1r+\frac4{r^2}.
\]
For $a=-2$, we have instead
\[
\frac{r+(-2+\lambda)^2}{(r+4-\lambda^2)^2} \leq\frac1{r+4}.
\]
Hence, recall that
\[
G_0\Psi_{-2}=-(2+\lambda)^{-1}\Psi_{-2}.
\]
we then obtain by using \eqref{ang:Xi} that 
\[
\begin{aligned}
\|G_0\Xi\|^2
&\leq
\frac{2\rho^2}{9(2+\lambda)^2}
+\frac{4v}{9}\left(\frac1h+\frac4{h^2}\right)
+\frac{2v}{9(h+4)} \\
&\leq
\frac{\rho^2}{18}
+\frac{4(1-\rho^2)}9(h^{-1}+4h^{-2})
+\frac{2(1-\rho^2)}{9(h+4)}
=M(\rho,h).
\end{aligned}
\]
This proves \eqref{ang:squared-loss}.

{\color{blue}Step 4.} Finally, we prove \eqref{ang:perturbation-loss}. Since $\|kG_0W_c\|\leq k/\delta<1$, the resolvent identity gives
\[
G_k-G_0 =-kG_0W_c(I+kG_0W_c)^{-1}G_0.
\]
Consequently,
\[
\begin{aligned}
\bigl|\langle\Xi,(G_k-G_0)\Xi\rangle\bigr|
&\leq k\|W_c\|\, \|(I+kG_0W_c)^{-1}\|\, \|G_0\Xi\|^2
\leq\frac{k}{1-k/\delta}M(\rho,h).
\end{aligned}
\]
This proves \eqref{ang:perturbation-loss}.
\end{proof}

\subsection{Proof of the angular criterion} \label{sec:proof_angular_criteria}

\begin{proof}[Proof of Proposition \ref{ang:criterion}]

{\color{blue}Step 1.} We first establish the Schur complement of the concerned operator $H^{(3)}_{k,q,1/2}-\lambda$ adapted to the decomposition
\[
L^2(\Omega;\mathbb C^8)
=\mathbb C\Psi_1\oplus\mathscr H_c,
\]
by which we will apply the results obtained in Section \ref{sec:complement_resolvent} to prove the desired gap estimate. Specifically, we have
\[
H^{(3)}_{k,q,1/2}-\lambda
=
\begin{pmatrix}
1-\lambda+\frac{2k\rho}{3}
&k\langle\Xi,\cdot\rangle\\
k\Xi&H_{0,c}+kW_c-\lambda
\end{pmatrix}.
\]
Its domain is $\mathbb C\Psi_1\oplus \bigl(\operatorname{Dom}(H_0)\cap\mathscr H_c\bigr)$, and the off-diagonal blocks are bounded. In particular, when $k<\delta$, the lower-right block is invertible. Thus, the full operator is invertible if and only if the scalar Schur complement is nonzero, i.e.,
\[
s_\lambda :=1-\lambda+\frac{2k\rho}{3} -k^2\langle\Xi,G_k\Xi\rangle \neq 0.
\]
By Lemma \ref{ang:resolvent},
\begin{equation}\label{ang:schur-lower}
s_\lambda\geq 1-\lambda+\frac{2k\rho}{3} -\frac{8k^2(1-\rho^2)}{9h} -\frac{k^3}{1-k/\delta}M(\rho,h).
\end{equation}
Therefore, the sufficient conditions
\begin{equation}\label{ang:rho-criterion}
k<\delta,
\quad
\frac{2k\rho}{3}>
\frac{8k^2(1-\rho^2)}{9h}
+\frac{k^3}{1-k/\delta}M(\rho,h)
\end{equation}
exclude every $\lambda\in[0,1]$ from the spectrum.

{\color{blue}Step 2.} We now verify $k<\delta$ throughout the stated parameter range. Note that
\[
\begin{aligned}
0<h=q^2+2k(1+\sqrt{1-q^2})
=(1+k)^2-(\sqrt{1-q^2}-k)^2
\leq\frac{49}{25}<2.
\end{aligned}
\]
Consequently,
\[
\delta=\sqrt{1+h}-1=d_0,
\]
where $d_0$ is the same as in \eqref{ang:parameters}. Moreover,
\[
d_0>k
\quad\Longleftrightarrow\quad
q^2+2k\sqrt{1-q^2}>k^2.
\]
The function $t\mapsto t+2k\sqrt{1-t}$ is concave on $[0,1]$, and hence
\[
q^2+2k\sqrt{1-q^2} \geq\min\{2k,1\}>k^2.
\]
Thus $d_0>k$, which proves $k<\delta$.

{\color{blue}Step 3.} Now we prove that, if $F(k,q)>0$, then
\begin{equation} \label{eq:half_gap_estimate}
\operatorname{spec}(H^{(3)}_{k,q,1/2}) \cap[0,1]=\varnothing.
\end{equation}
This is proved by showing that the scalar Schur complement $s_{\lambda}>0$. Indeed, for $0<h<2$, the coefficient of $r^2$ in $M(r,h)$ is
\[
\frac1{18}-\frac4{9h}-\frac{16}{9h^2}-\frac2{9(h+4)}<0.
\]
Thus $M(r,h)$ is non-increasing for $0\leq r\leq1$. By Lemma \ref{rad:rho-bound},
\[
\rho\geq\rho_0,\quad
1-\rho^2\leq1-\rho_0^2,\quad
M(\rho,h)\leq M(\rho_0,h).
\]
Substituting these inequalities into \eqref{ang:schur-lower}, and using $\delta=d_0$, gives
\[
\begin{aligned}
s_\lambda
&\geq
1-\lambda+\frac{2k\rho_0}{3}
-\frac{8k^2(1-\rho_0^2)}{9h}
-\frac{k^3}{1-k/d_0}M(\rho_0,h)\\
&=1-\lambda+k^2F(k,q),
\end{aligned}
\]
where we used the identity $\frac{2\rho_0}{3k}
=\frac4{3(2k+q^2)}$. This proves \eqref{eq:half_gap_estimate}.

{\color{blue}Step 4.} Finally we prove the full estimate in Proposition \ref{ang:criterion} by exploiting the symmetry of $H^{(3)}_{k,q,1/2}$. Indeed, since the unitary matrix $\zeta_2$ anti-commutes with every term of $H^{(3)}_{k,q,1/2}$, i.e., $\zeta_2H^{(3)}_{k,q,1/2}\zeta_2
=-H^{(3)}_{k,q,1/2}$, we know the spectrum of $H^{(3)}_{k,q,1/2}$ is symmetric about zero. Then the proof of \eqref{ang:gap} is completed by \eqref{eq:half_gap_estimate}.
\end{proof}

\section{Completion of the small-separation estimate} \label{sec:bridge}

In this section, we show that the auxiliary operator $H^{(3)}_{k,q,1/2}$ has the desired spectral gap $[-1,1]$ for all choices of $k>0$ and $q\in [0,1]$:
\begin{proposition}\label{bridge:gap}
For every $k>0$ and $q\in[0,1]$,
\begin{equation}\label{bridge:full-gap}
\operatorname{spec}(H^{(3)}_{k,q,1/2})\cap[-1,1] =\varnothing.
\end{equation}
\end{proposition}
To this end, we first point out that the angular criterion from Section~\ref{sec:angular} covers all $q\in[0,1]$ when $k$ is small. Specifically, the following estimate, together with Proposition \ref{ang:criterion}, proves Proposition \ref{bridge:gap} for $0<k\leq3/16$:
\begin{lemma}\label{bridge:small-k}
The function $F$ defined in \eqref{ang:F} satisfies
\[
 F(k,q)>0
 \quad\text{for }0<k\leq3/16,\quad 0\leq q\leq1.
\]
\end{lemma}
The proof of Lemma \ref{bridge:small-k} involves computer-assisted (yet exact) calculation, which is presented in Appendix~\ref{sec:finite_certificate_small_k}. On the remaining compact rectangle, i.e. $3/16\leq k\leq 2/5$ (recall that we have proved the desired gap estimate in Corollary \ref{end:large-k} for $k\geq 2/5$), the gap estimate is obtained by combining Proposition \ref{ang:criterion} with the endpoint estimate \eqref{end:chord}. To see the idea, recall the notions
\[
s_k=(1+2k)e^{-2k},\quad c_k=\frac{\sqrt{8+s_k^2}+s_k}{4}.
\]
For $0\leq k\leq2/5$, the elementary bounds
\begin{equation}\label{bridge:ck-polynomial}
 s_k\leq1-k^2, \quad c_k\leq1-\frac{k^2}{4}
\end{equation}
are proved in Appendix~\ref{sec:finite_certificate_prep}. Consequently, \eqref{end:chord} gives
\[
\nu_{k,1/2}(q)-1\geq C(k,q),\quad 0<k\leq2/5,\quad 0\leq q\leq1,
\]
where
\begin{equation}\label{bridge:C}
 C(k,q):=
 q^2+\left[1-\left(1-\frac{k^2}{4}\right)q\right]
       \left(1+\frac43k\right)-1,
\end{equation}
Thus, either $F(k,q)>0$ or $C(k,q)>0$ is sufficient for \eqref{bridge:full-gap}. This is indeed true, as proved in Appendix~\ref{sec:finite_certificate_separation}:
\begin{lemma}\label{bridge:compact}
For any $(k,q)\in [3/16,2/5]\times[0,1]$, we have either $F(k,q)>0$ or $C(k,q)>0$.
\end{lemma}
We briefly sketch the idea of proof: letting
\[
s=s(q):=\sqrt{1-q^2},\quad \ell(k):=\frac95k-\frac9{40},
\]
then, on $[3/16,2/5]\times[0,1]$, we have
\begin{equation}\label{bridge:classical-separator}
\begin{split}
s(q)\leq\ell(k)&\ \Longrightarrow\ C(k,q)>\frac1{800},\\
s(q)\geq\ell(k)&\ \Longrightarrow\ F(k,q)>0.
\end{split}
\end{equation}
This displays the role of the two estimates: the endpoint criterion covers the side near $q=1$, and the angular criterion covers the other side. Both hold strictly on the separating curve, so no boundary case is left open. The polynomial expansions and coefficient comparisons that establish these implications are given in Appendix~\ref{sec:finite_certificate_separation}.

\begin{proof}[Proof of Proposition~\ref{bridge:gap}]
Fix $k>0$ and $q\in[0,1]$. If $0<k\leq3/16$, Lemma~\ref{bridge:small-k} gives $F(k,q)>0$, and Proposition~\ref{ang:criterion} yields the desired gap.

If $3/16\leq k\leq2/5$, Lemma~\ref{bridge:compact} gives either $F(k,q)>0$ or $C(k,q)>0$. The first condition again implies the gap by Proposition~\ref{ang:criterion}. In the second case, \eqref{end:chord} and \eqref{bridge:ck-polynomial} give $\nu_{k,1/2}(q)-1\geq C(k,q)>0$, which implies the same conclusion.

Finally, if $k\geq2/5$, the result follows from Corollary~\ref{end:large-k}. The proof then completes.
\end{proof}

\section{Proof of Theorems \ref{thm:main} and \ref{thm:cross}}
\label{sec:conclusion}

In this section, we first complete the proof of Theorem \ref{thm:cross}, which follows directly by the results established in Sections \ref{sec:radial}-\ref{sec:bridge}, especially Proposition \ref{bridge:gap}. For the proof of Theorem \ref{thm:main}, we first compare the Birman-Schwinger operator of a finite Coulomb configuration with a finite matrix of block norms. This comparison gives a strict contraction at the atomic energy, using Theorem \ref{thm:cross}, from which the concentration inequality and its equality case follow.

\begin{proof}[Proof of Theorem~\ref{thm:cross}]
Fix $m>0$, $0\leq E<m$, and distinct centers $A,B$. Recall the notation
\[
 \kappa=\sqrt{m^2-E^2},\quad
 k=\frac{\kappa|A-B|}{2}>0,\quad
 T_E=w_B R_Ew_A.
\]
Suppose that \eqref{intro:cross} fails. Since $T_E$ is compact, it has a singular value $s\geq m/\kappa$. Moreover, $T_E$ preserves the axial angular momentum (see Proposition \ref{ax:singular-vector}), so a corresponding singular pair can be chosen in one sector $j\in\mathbb Z+\tfrac12$. Hence, by Proposition~\ref{ax:singular-vector}, there is a nonzero vector $\Phi$ satisfying
\[
 H^{(3)}_{k,q,j}\Phi=\lambda\Phi,\quad
 \lambda=\frac{m}{s\kappa}\in(0,1],\quad
 q=\frac Em\lambda\in[0,1].
\]
If $|j|\geq3/2$, this contradicts
\eqref{ax:higher-sectors}; hence it must hold that $j\in\{\pm 1/2\}$. But then we draw a contradiction to Proposition~\ref{bridge:gap} by recalling the fact that $H^{(3)}_{k,q,1/2}$ is unitarily equivalent to $H^{(3)}_{k,q,-1/2}$. This proves Theorem~\ref{thm:cross}.
\end{proof}

\begin{proof}[Proof of Theorem~\ref{thm:main}]
The proof proceeds in four steps. We first establish a strict contraction for the Birman-Schwinger operator at the atomic energy. We then derive the corresponding resolvent formula and use it to exclude the lower part of the gap. Finally, we prove the existence of a first gap eigenvalue above the atomic energy.

{\color{blue}Step 1.} We first compare the Birman-Schwinger operator associated with $H_{V}$ with a finite matrix of size $N\times N$. To this end, we define
\begin{equation}\label{eq:K_E_def}
\begin{gathered}
E_G:=m\sqrt{1-G^2},\qquad
W\psi:=(\sqrt{g_i}\,w_{A_i}\psi)_{i=1}^N,\\
\mathcal K_E:=WR_EW^*\qquad(-m<E<m),
\end{gathered}
\end{equation}
where $R_E=(D_m-E)^{-1}$ is the free resolvent and $w_{A_i}(x)=|x-A_i|^{-1/2}$ is the potential weight. The operator $\mathcal K_E$ is bounded and self-adjoint on $\mathcal H^N$. Its $(i,j)$ block is
\[
 (\mathcal K_E)_{ij} =\sqrt{g_i g_j}\,w_{A_i}R_Ew_{A_j}.
\]
Since $E_{G}<m$, $R_{E_G}$ is well-defined. Then we introduce the symmetric matrix $B$ by the block norms
\[
 B_{ij}:=\sqrt{g_i g_j}\,
 \bigl\|w_{A_i}R_{E_G}w_{A_j}\bigr\|.
\]
The one-center identity \eqref{intro:atomic-norm} and Theorem~\ref{thm:cross} yield a complete estimate of all entries of the matrix $B$:
\begin{equation}\label{con:matrix}
 B_{ii}=\frac{g_i}{G},\qquad
 0\leq B_{ij}<\frac{\sqrt{g_i g_j}}{G}
 \quad(i\ne j).
\end{equation}
Importantly, the norm of $B$ dominates that of the Birman-Schwinger operator:
\begin{equation}\label{con:matrix_1}
 \|\mathcal K_{E_G}\|\leq\|B\|.
\end{equation}
It follows by noticing that, for $f=(f_1,\ldots,f_N)\in\mathcal H^N$ and
$u_i:=\|f_i\|$,
\[
 \|(\mathcal K_{E_G}f)_i\|
 \leq\sum_{j=1}^N B_{ij}u_j.
\]
We also claim that
\begin{equation} \label{con:contraction}
\rho:=\|B\|<1 .
\end{equation}
To this end, define $v=(v_1,\ldots,v_N)$ with $ v_i:=\sqrt{\frac{g_i}{G}}$; then $\|v\|=1$ and all components $v_i$ are strictly positive. For any unit vector $z\in\mathbb C^N$,
\[
 |z^*Bz|
 \leq |z|^{\mathsf T}B|z|
 \leq (v\cdot|z|)^2
 \leq1,
\]
with the last equality holding if and only if $|z|=v$. However, in that case, the second inequality is strict due to the estimates \eqref{con:matrix}, since $N\geq2$ and all components of $v$ are positive. Consequently,
$|z^*Bz|<1$ for every unit vector $z$, which proves \eqref{con:contraction} by noting that $B$ is symmetric.

{\color{blue}Step 2.} We now write the resolvent of $H_V$ in a standard Birman-Schwinger form using the operator $\mathcal K_E$ and $W$ introduced in Step 1. To be precise, we claim that whenever $I-\mathcal K_E$ is invertible,
\begin{equation}\label{con:resolvent}
(H_V-E)^{-1}=R_E+R_EW^*(I-\mathcal K_E)^{-1}WR_E=: \mathcal S_E.
\end{equation}
Here, the domains of the involved operators are specified as follows: 1) $W:H^{1/2}\longrightarrow\mathcal H^N$, $W^*:\mathcal H^N\longrightarrow H^{-1/2}$ are bounded by the fractional Hardy inequality, and 2) $R_E:H^{-1/2}\to H^{1/2}$ is bounded by the standard regularity theory of the free Dirac operator. Here and below, the Sobolev spaces are those of $\mathbb C^4$-valued functions on $\mathbb R^3$. Moreover, since the multiplication by $w_A$ is compact from $H^1$ into $L^2$, it follows that the compositions
\[
 WR_E:\mathcal H\longrightarrow\mathcal H^N,
 \quad
 R_EW^*:\mathcal H^N\longrightarrow\mathcal H
\]
are compact.

To prove \eqref{con:resolvent}, set $\psi:=\mathcal S_Ef$ for $f\in\mathcal H$. The preceding mapping properties give $\psi\in H^{1/2}$, and by direct calculation,
\begin{align*}
 W\psi=WR_Ef+\mathcal K_E(I-\mathcal K_E)^{-1}WR_Ef=(I-\mathcal K_E)^{-1}WR_Ef.
\end{align*}
Using $W^*W=V$ in the distributional sense (recall the definition \eqref{intro:potential} of the potential $V$), we obtain
\[
 (D_m-V-E)\psi=f.
\]
Thus, $\psi\in\Dom(H_V)$ by \eqref{pre:domain}, and
$(H_V-E)\psi=f$. This also proves that $H_V-E$ is surjective. Since $E$ is real and $H_V$ is self-adjoint, we know $\ker(H_V-E)=\operatorname{Ran}(H_V-E)^\perp=\{0\}$, and hence $H_V-E$ is bijective. This completes the proof of $\mathcal S_E=(H_V-E)^{-1}$.

{\color{blue}Step 3.} We next exclude the interval $(-m,E_G]$ from the spectrum of $H_V$. First, for $-m<E<E_G$, the resolvent identity gives the monotonicity of free resolvents:
\begin{equation}\label{con:order}
 R_{E_G}-R_E =(E_G-E)R_{E_G}R_E\geq0.
\end{equation}
This is proved by noting that, in terms of spectral calculus, both resolvents are functions of the free operator $D_m$, and their scalar values have the same sign on each spectral component of $D_m$. Importantly, multiplying the two sides of \eqref{con:order} by the (positive) potential weights, we obtain
\[
 \mathcal K_E
 \leq\mathcal K_{E_G}\leq\rho I,
\]
where the second inequality follows by \eqref{con:matrix_1}. Together with \eqref{con:contraction}, we know
\[
I-\mathcal K_E\geq(1-\rho)I>0 \quad(-m<E\leq E_G).
\]
Therefore, $I-\mathcal K_E$ is invertible for $E\in(-m,E_G]$. The resolvent formula \eqref{con:resolvent} then implies
\begin{equation}\label{con:no-lower-spectrum}
 \operatorname{spec}(H_V)\cap(-m,E_G]=\varnothing.
\end{equation}

{\color{blue}Step 4.} It remains to prove that the first in-gap eigenvalue of $H_{V}$ exists and lies strictly above $E_G$. By \eqref{con:resolvent}, the difference $\mathcal S_{E_G}-R_{E_G}$ is compact. Hence by the invariance of essential spectrum under compact resolvent perturbation (cf. \cite[Chapter IV, Problem 5.38]{Kato1995})
\[
\operatorname{spec}_{\mathrm{ess}}(H_V)=\operatorname{spec}_{\mathrm{ess}}(D_{m})=(-\infty,-m]\cup[m,\infty).
\]
Moreover, for $E_G<E<m$, the resolvent identity gives
\begin{equation}\label{con:compact-difference}
 \mathcal K_E-\mathcal K_{E_G} =(E-E_G)(WR_E)(R_{E_G}W^*).
\end{equation}
The right-hand side is compact and depends continuously on $E$ in operator norm on compact subintervals of the gap. Consequently, $\mathcal K_E$ is norm-continuous in $E$, and
\[
 \operatorname{spec}_{\mathrm{ess}}(\mathcal K_E)
 \subset[-\rho,\rho].
\]

Recall that, for each single-center potential with the center $i$ and $0<E<m$, the positive atomic endpoint in \eqref{intro:atomic-norm} is attained, as shown just before \eqref{pre:atomic-energy}. Hence, by inserting a corresponding normalized vector into the $i$th component of $\mathcal H^N$, with all other components equal to zero, the definition \eqref{eq:K_E_def} gives
\[
 \sup\operatorname{spec}(\mathcal K_E)
 \geq\frac{g_i m}{\sqrt{m^2-E^2}}
 \longrightarrow\infty
 \qquad(E\uparrow m).
\]
Hence, we can define
\[
 \Lambda(E):=\sup\operatorname{spec}(\mathcal K_E),
 \quad
 E_1:=\inf\{E\in(E_G,m):\Lambda(E)\geq1\}.
\]
We claim that $E_1$ is the first eigenvalue of $H_{V}$ that satisfies the desired estimate. First, $E_1\in(E_G,m)$ follows by the definition of $E_1$ and the fact $\sup\operatorname{spec}(\mathcal K_{E_G})\leq\rho<1$ (recall \eqref{con:matrix_1}-\eqref{con:contraction}). To show that $E_1$ is indeed an eigenvalue of $H_{V}$, we show that $1$ is an eigenvalue of $\mathcal K_{E_1}$; then the standard Birman-Schwinger correspondence concludes the proof. In fact, we note that $\Lambda(E_1)=1$ by the norm continuity of $\mathcal K_{E}$. Moreover, since the essential spectrum of $\mathcal K_{E_1}$ lies below $1$, there is a nonzero $f$ with $\mathcal K_{E_1}f=f$. Then $\psi:=R_{E_1}W^*f\in H^{1/2}$ satisfies $W\psi=f\ne0$ and $(D_m-E_1)\psi=W^*W\psi=V\psi$. Thus, $H_V\psi=E_1\psi$ by \eqref{pre:domain}. Finally, for $E_G<E<E_1$, the definition of $E_1$ gives $\Lambda(E)<1$, so $I-\mathcal K_E$ is invertible. This concludes that $H_{V}$ does not have an eigenvalue in $(E_G,E_1)$ by \eqref{con:resolvent} and completes the proof.
\end{proof}

\begin{proof}[Proof of Corollary~\ref{cor:general-charges}]
Fix $E_G=m\sqrt{1-G^2}$ as in \eqref{eq:K_E_def}. If $\mu=G\delta_A$, the first positive eigenvalue \eqref{pre:atomic-energy} for the atomic case satisfies $\lambda_1(H_\mu)=E_G$. We note that the atomic min-max formula holds also for the critical distinguished realization at $G=1$; see \cite[Theorem~2.2]{ELScritical2019}.

Suppose now that $\mu$ is not concentrated at one point. Define
\[
 \mathcal H_\mu:=L^2(\R^3,d\mu;\mathcal H),\quad
 (W_\mu\psi)(A):=w_A\psi,\quad
 \mathcal K_{\mu,E_G}:=W_\mu R_{E_G}W_\mu^* .
\]
Note that the fractional Hardy inequality, which is uniform in $A$, gives a bounded map $W_\mu:H^{1/2}\to\mathcal H_\mu$, and $W_\mu^*W_\mu=V_\mu$ in Sobolev duality. For $b(A,B):=\|w_A R_{E_G}w_B\|$, we know that, since translations act strongly continuously on $H^{1/2}$, $\mathcal H$ and $H^{-1/2}$, the operator inside this norm is strongly continuous in $(A,B)$ and hence $b$ is Borel. On simple $\mathcal H$-valued functions the kernel of $\mathcal K_{\mu,E_G}$ in the duality pairing is $w_A R_{E_G}w_B$. Thus, by the Cauchy-Schwarz inequality, it holds that
\begin{equation} \label{eq:general-charge_proof_1}
\|\mathcal K_{\mu,E_G}\|^2
 \leq \iint b(A,B)^2\,d\mu(A)\,d\mu(B)<1.
\end{equation}
This is justified by noting that 1) \eqref{intro:atomic-norm} and Theorem~\ref{thm:cross} give $b(A,B)\leq 1/G$, with strict inequality for $A\ne B$, and 2) the product measure has mass $G^2$.

Now we apply the estimate \eqref{eq:general-charge_proof_1} to show that $E_{G}\notin \sigma(H_{t\mu})$ for any $0<t\leq1$. First, since every atom of $t\mu$ has mass strictly below one, we know $\Dom(H_{t\mu})\subset H^{1/2}$ by \cite[Theorem~2.1]{ELSI}. If $H_{t\mu}\psi=E_G\psi$ for $\psi\ne0$, then $(D_m-E_G)\psi=tW_\mu^*W_\mu\psi$. Consequently, $W_\mu\psi=t\mathcal K_{\mu,E_G}W_\mu\psi$, with $W_\mu\psi\ne0$ since $D_m-E_G$ is invertible. This contradicts to the fact that $\|K_{\mu,E_G}\|<1$ proved in \eqref{eq:general-charge_proof_1}, and hence justifies the claim.

Finally, we note that by \cite[Section~2.2]{ELSII}, after scaling to mass $m$, the first min-max level of $H_{t\mu}$ is continuous in $t$, tends to $m$ as $t\downarrow0$, and equals the lowest gap eigenvalue whenever it lies above $-m$ and $t>0$. Since $-m<E_G<m$, reaching $E_G$ would give an eigenvalue there. Hence, by the results obtained above, we conclude that the first energy level stays strictly above $E_G$ up to $t=1$. This proves strictness for every such $\mu$ and identifies the eigenvalue of Theorem~\ref{thm:main} with the first min-max level as well.
\end{proof}

\begin{remark}
Note that Theorem~\ref{thm:main} compares a finite Coulomb configuration with the single atom carrying its total charge. It does not assert that the first gap eigenvalue decreases under every contraction of the nuclear positions. Such monotonicity, conjectured in \cite{ELSsurvey}, would imply the non-strict concentration bound by contracting all centers to a single point.
\end{remark}

\bigskip
\textbf{Conflict of Interest Statement}: On behalf of all authors, the
corresponding author states that there is no conflict of interest.

\bigskip
\textbf{Data Availability Statement}: On behalf of all authors, the
corresponding author confirms that the current paper does not have
associated data.

\bigskip
\textbf{AI Usage}: The authors developed the proof strategy, including the Birman-Schwinger formalism and the proposed two-center estimate. The details are worked out assisted by ChatGPT 6 Astra, particularly the analysis in the small-separation regime. The whole manuscript is substantially improved and rewritten by the authors. In particular, the authors have isolated the technical calculation of computer-generated exact numerics (e.g., Lemma \ref{bridge:small-k} and \ref{bridge:compact}) from the main text to ensure the readers' smooth understanding of the main idea of proof. The authors take full responsibility for the content of this paper.

\appendix

\section{Exact scalar inequalities} \label{app:scalar}

This appendix contains the scalar verifications used in Sections~\ref{end:section} and~\ref{sec:bridge}. They concern only explicit functions of $k$ and $q$. After elementary bounds on radicals and the exponential, each verification reduces to a polynomial identity and finitely many rational inequalities that can be checked both by hand or, more directly, by software like Mathematica.

\subsection{The large-separation regime}\label{app:large-scalar}

\begin{proof}[Proof of Corollary \ref{end:large-k}]
Since $c_k<1$, the estimate \eqref{end:chord} applies throughout $0\leq q\leq1$. Before applying \eqref{end:chord}, we first record an explicit bound for $c_k$ and $A_k$. By the definition \eqref{end:constants}, it is direct to check that $c_k$ decreases with $k$ while $A_{k}$ increases; hence, for $k\geq \frac{2}{5}$, a direct calculation shows that
\begin{equation*}
c_k<\frac{47}{50},\quad A_k=1+\frac{4k}{3}\geq\frac{23}{15}.
\end{equation*}
Thus, for $0\leq q\leq1$, we know that
\begin{equation*}
\nu_{k,j}(q)-1
\geq q^2+A_k(1-c_kq)-1
\geq q^2+\frac{23}{15}
                  \left(1-\frac{47}{50}q\right)-1 >0 .
\end{equation*}
This proves \eqref{end:large-k-bound}.
\end{proof}

\subsection{A polynomial form of the endpoint criterion}\label{sec:finite_certificate_prep}

\begin{proof}[Proof of \eqref{bridge:ck-polynomial}]
Indeed, differentiating $s_k$ gives $1-s_k=4\int_0^k te^{-2t}\,dt$ and using the Taylor expansion $e^{-2t}\geq1-2t+2t^2-\frac43t^3$ ($t\geq 0$), we obtain
\begin{align*}
 1-s_k \geq k^2\left( 2-\frac83k+2k^2-\frac{16}{15}k^3 \right).
\end{align*}
It is clear that the polynomial in parentheses is decreasing on $[0,2/5]$. Its value at $k=2/5$ is greater than one. Therefore,
\[
 1-s_k\geq k^2,
\]
which proves the first inequality in \eqref{bridge:ck-polynomial}. On the other hand, since $0\leq s_k\leq1$, we have
\[
 c_k \leq\frac{3+s_k}{4} \leq 1-\frac{k^2}{4},
\]
which proves the second inequality.
\end{proof}

\subsection{Positivity of the angular criterion for small $k$}\label{sec:finite_certificate_small_k}

\begin{proof}[Proof of Lemma~\ref{bridge:small-k}]
The idea is to bound the radicals appearing in $F$ (see \eqref{ang:F}) by rational expressions. Then we will bound the denominators appearing in the rational expressions from below, and prove the positivity of numerator by expanding it in the Bernstein basis.

{\color{blue}Step 1.} Let us recall the definition of $F(k,q)$ from \eqref{ang:F}:
\begin{equation} \label{ang:F_recall}
F(k,q)=\frac4{3(2k+q^2)}
-\frac{8(1-\rho_0^2)}{9\mu_*}
-\frac{k}{1-k/d_0}M(\rho_0,\mu_*).
\end{equation}
We first bound the denominators $\mu_*$ and $1-k/d_0$ from below. To this end, define the auxiliary scalars
\[
 B_0:=q^2+2k,\quad w:=q^2(q^2+4k),
\]
by which we have
\begin{equation} \label{eq:auxiliary_scalars_for_F}
\rho_0=\frac{2k}{B_0}, \quad 1-\rho_0^2=\frac{w}{B_0^2}.
\end{equation}
We also write
\[
 m_0:=(1-2k)q^2+4k,\quad D_0:=(2-k)m_0-4k.
\]
Since $\sqrt{1-q^2}\geq 1-q^2$, the definition of $\mu_*$ in \eqref{rad:gap} gives
\[
\mu_*=q^2+2k\bigl(1+\sqrt{1-q^2}\bigr) \geq q^2+2k(2-q^2) =m_0>0.
\]
On the other hand, using $\sqrt{1+x}\leq1+x/2$ for $x\geq0$, we obtain
\[
d_0 =\frac{\mu_*}{\sqrt{1+\mu_*}+1} \geq\frac{2\mu_*}{4+\mu_*} \geq\frac{2m_0}{4+m_0},
\]
where the last inequality follows from the monotonicity of
$x\mapsto2x/(4+x)$. Hence, by noting that
\[
 D_0=(2-k)(1-2k)q^2+4k(1-k)>0,
\]
we obtain
\begin{equation*}
1-\frac{k}{d_0} \geq1-\frac{k(4+m_0)}{2m_0} =\frac{D_0}{2m_0}>0.
\end{equation*}
In conclusion, using \eqref{eq:auxiliary_scalars_for_F} and the two estimates of denominators obtained above, we know
\begin{equation} \label{eq:Fkq_rational_gound}
F(k,q)\geq \frac4{3B_0} -\frac{8w}{9B_0^2m_0} -\frac{2km_0}{D_0} M\left(\frac{2k}{B_0},m_0\right),
\end{equation}
where we have also applied the fact that the function $h\mapsto M(r,h)$ defined in \eqref{ang:M} is nonnegative and non-increasing, for each fixed $r\in[0,1]$. Substituting the explicit expression for $M$ and collecting the terms over a common denominator gives
\begin{equation}\label{bridge:polynomial-lower}
 F(k,q)\geq
 \frac{4N(k,q^2)}{9B_0^2m_0(m_0+4)D_0},
\end{equation}
where the auxiliary polynomial $N$ is defined by
\begin{align}
 N(k,q^2):={}&
 3B_0m_0(m_0+4)D_0-2w(m_0+4)D_0
 \nonumber\\
 &-k^3m_0^2(m_0+4)
 -2kw(m_0+4)^2
 -kwm_0^2.
 \label{bridge:N}
\end{align}
All factors in the denominator of
\eqref{bridge:polynomial-lower} are strictly positive.
Thus, it remains to prove that
\[
 N(k,q^2)>0
 \quad\text{for }\, 0<k\leq3/16,\quad 0\leq q\leq1.
\]

{\color{blue}Step 2.} To prove the positivity of the polynomial $N(k,q^2)$, we exploit its Bernstein expansions. Rescale $t:=16k/3$ so that $0<t\leq 1$, and denote the Bernstein polynomials by
\[
 b_{i,n}(t):=\binom ni t^i(1-t)^{n-i}, \quad 0\leq i\leq n.
\]
In particular, the polynomial in \eqref{bridge:N} admits the expansion
\begin{equation}\label{bridge:Bernstein}
 N(3t/16,q^2)
 =\sum_{i=0}^{6}\sum_{j=0}^{4}
 c_{ij}b_{i,6}(t)b_{j,4}(q^2),
\end{equation}
where the coefficient matrix $(c_{ij})$ is given by
\begingroup
\setlength{\arraycolsep}{6pt}
\renewcommand{\arraystretch}{1.8}
\[
\begin{pmatrix}
0&0&0&2&10\\
0&0&\tfrac{1}{3}&\tfrac{83}{32}&\tfrac{77}{8}\\
0&\tfrac{21}{160}&\tfrac{19}{24}&\tfrac{7971}{2560}&\tfrac{713}{80}\\
\tfrac{81}{640}&\tfrac{1197}{2560}&\tfrac{56293}{40960}&\tfrac{1147753}{327680}&\tfrac{637157}{81920}\\
\tfrac{81}{160}&\tfrac{21771}{20480}&\tfrac{1010203}{491520}&\tfrac{2411737}{655360}&\tfrac{1002247}{163840}\\
\tfrac{40905}{32768}&\tfrac{1017285}{524288}&\tfrac{2182655}{786432}&\tfrac{7416933}{2097152}&\tfrac{1995053}{524288}\\
\tfrac{635607}{262144}&\tfrac{6453819}{2097152}&\tfrac{7201957}{2097152}&\tfrac{24577225}{8388608}&\tfrac{1449503}{2097152}
\end{pmatrix}.
\]
\endgroup
Indeed, the identity~\eqref{bridge:Bernstein} can be checked by direct expansion. Alternatively, since both sides of \eqref{bridge:Bernstein} have degree at most six in $t$ and four in $q^2$, it suffices to check their equality at the $35$ rational points
\[
 (t,q^2)=(i/6,j/4),
 \quad 0\leq i\leq6,\, 0\leq j\leq 4,
\]
which can be verified by direct calculation (as checked by the authors using Mathematica).

We now apply \eqref{bridge:Bernstein} to prove that the numerator in \eqref{bridge:polynomial-lower} is strictly positive. Since 1) all entries of $(c_{ij})$ are non-negative, 2) every entry in its row $i=3$ is strictly positive, and 3) for $0<t<1$, we have $b_{3,6}(t)>0$, while
\[
 b_{j,4}(a)\geq0, \quad \sum_{j=0}^{4}b_{j,4}(a)=1 \quad (0\leq a\leq1),
\]
we know
\[
N(3t/16,q^2) \geq b_{3,6}(t) \sum_{j=0}^{4}c_{3j}b_{j,4}(q^2)>0.
\]
At $t=1$, only the row $i=6$ contributes, and all entries in that row are strictly positive. Hence,
\[
 N(3/16,q^2)=\sum_{j=0}^{4}c_{6j}b_{j,4}(q^2)>0 \quad (0\leq q\leq1).
\]
This concludes the proof.
\end{proof}

\subsection{Verification on the remaining compact rectangle}\label{sec:finite_certificate_separation}

\begin{proof}[Proof of Lemma~\ref{bridge:compact}]

{\color{blue}Step 1.} We prove the first implication in \eqref{bridge:classical-separator}. For ease of notation, we write
\begin{equation*}
C(k,q)=q^2+4k/3-Bq,\quad
\text{ with }\, A:=\frac{k^2}{4}+\frac{k^3}{3},\, B:=1+\frac43k-A .
\end{equation*}
For $k\in [3/16,2/5]$, it holds that $0<\ell(k)<3/5$ and $0<B\leq23/15$. If $s\leq \ell(k)$, then $q\geq\sqrt{1-\ell(k)^2}>4/5$, and hence $\partial_q C=2q-B>1/15$. Therefore, it suffices to calculate the end-point value $C(k,\sqrt{1-\ell(k)^2})$. To this end, recall the elementary inequality
\[
 \sqrt{1-z}\leq1-\frac z2-\frac{z^2}{8},\quad 0\leq z\leq1,
\]
by which one can check that
\[
C\bigl(k,\sqrt{1-\ell(k)^2}\bigr)
\geq E(k):=A+\left(\frac B2-1\right)\ell(k)^2+\frac B8\ell(k)^4.
\]
Here $E$ is a polynomial of degree seven. We expand $E$ in terms of the difference $u:=k-5/16$ ($-1/8\leq u\leq 7/80$): writing $E(k)=\sum_{i=0}^7 e_i u^i$, one can check that
\begin{align*}
 e_0&=\frac{6585859303}{4026531840000},&
 e_1&=-\frac{875274869}{83886080000},\\
 e_2&=\frac{3167587449}{5242880000},&
 e_3&=\frac{660652153}{196608000},\\
 e_4&=\frac{6532731}{4096000},&
 e_5&=\frac{295137}{1280000},\\
 e_6&=-\frac{85293}{80000},&
 e_7&=-\frac{2187}{5000}.
\end{align*}
The terms of degree at least three are estimated as
\[
\left|\sum_{i=3}^7e_i u^i\right| 
 \leq\frac12u^2.
\]
It follows that
\begin{align*}
 E(k)>\frac1{625}-\frac{11}{1000}|u|+\frac{u^2}{10}
 >\frac1{800},
\end{align*}
where we have applied the fact that $|u|\leq 1/8$. This proves the first implication in \eqref{bridge:classical-separator}.

{\color{blue}Step 2.} We next prove the second implication in \eqref{bridge:classical-separator}. Suppose that $s(q)\geq\ell(k)$, and write
\[
\begin{gathered}
b=1-s^2+2k,\qquad h=(1+s)(1-s+2k),\\
W=(1-s^2)(1-s^2+4k),\qquad D=(30-9k)h-64k.
\end{gathered}
\]
Thus $h=\mu_*$, $\rho_0=2k/b$, and $1-\rho_0^2=W/b^2$. Using the elementary inequality
\[
 \sqrt{1+h}\leq\frac3{10}h+\frac{17}{15},
\]
we know that
\[
 d_0=\frac h{\sqrt{1+h}+1}\geq\delta:=\frac{30h}{9h+64}.
\]
Moreover, $h-4k=(1-s)(1+s-2k)\geq0$, so
\[
 D\geq k(56-36k)>0,\qquad \delta>k.
\]
Since $d/(d-k)$ decreases for $d>k$, substitution into the definition of $F$ in \eqref{ang:F} gives
\begin{equation}\label{bridge:classical-rational-F}
 F(k,q)\geq\frac{4P(k,s)}{9b^2h(h+4)D},
\end{equation}
where
\begin{equation}\label{bridge:classical-P}
P=(3bh-2W)D(h+4) -15k\bigl[k^2h^2(h+4)+W(3h^2+16h+32)\bigr].
\end{equation}
Since the denominator is clearly positive, it suffices to prove that $P>0$. This is again achieved by an explicit polynomial expansion. Since $P$ has a factor $1+s$ as checked by the definition \eqref{bridge:classical-P}, we instead consider $Q:=P/(1+s)$, and expand it in powers of $1-s$:
\[
 Q(k,s)=\sum_{i=0}^7 A_i(k)(1-s)^i,
\]
with
\begin{align*}
 A_0&=96k^3(k+1)(28-18k-5k^2),\\
 A_1&=32k^2(15k^4+39k^3-309k^2-80k+100),\\
 A_2&=-8k(15k^5-36k^4-1029k^3+1565k^2+578k-236),\\
 A_3&=-12(15k^5+168k^4-1044k^3+504k^2+172k-40),\\
 A_4&=2(63k^4-2004k^3+3884k^2-736k-120),\\
 A_5&=417k^3-2622k^2+2048k-240,\\
 A_6&=6(45k^2-104k+30),\quad A_7=6(9k-5).
\end{align*}
Set
\[
 x:=\frac{80k-15}{17},\quad
 w:=\frac{355-153x}{400}=1-\ell(k),\quad
 y:=\frac{s-\ell(k)}{1-\ell(k)}.
\]
Then $0\leq x,y\leq1$ and $1-s=w(1-y)$. The binomial theorem gives the identity
\begin{equation}\label{bridge:classical-binomial}
Q(k,s)=\sum_{j=0}^7\binom7j y^j(1-y)^{7-j}B_j(x).
\end{equation}
Here the polynomials $B_j$ are explicitly given by
\begin{equation}\label{bridge:classical-Bj}
 B_j(x)=\sum_{i=0}^{7-j}
 \frac{\binom{7-j}{i}}{\binom7i}\,
 A_i\!\left(\frac{15+17x}{80}\right)
 \left(\frac{355-153x}{400}\right)^i ,
\end{equation}
which follows by expanding each $(1-y)^i$ in the degree-seven Bernstein basis. To show that $Q>0$ using \eqref{bridge:classical-binomial}, we record the coefficient bounds for each $B_j$ as follows: writing $B_j(x)=\sum_{i=0}^8a_{ji}x^i$, with absent coefficients set equal to zero, the following table records the lower bounds for $a_{j0},\ldots,a_{j3}$ (the first four entries), the indices $i\geq4$ with $a_{ji}<0$ (the set $N_j$), and an upper bound for $-\sum_{i\in N_j}a_{ji}$ ($L_j$):
\[
\begin{array}{c|rrrr|c|c}
 j&a_{j0}\geq&a_{j1}\geq&a_{j2}\geq&a_{j3}\geq&N_j&L_j\\ \hline
 0&97/2&-257/2&97/2&54&\{5,6,7,8\}&1\\
 1&75&-88&52&0&\{5,6,7,8\}&1\\
 2&85&-40&47&0&\{6,7,8\}&1\\
 3&79&0&0&0&\{7\}&1\\
 4&62&0&0&0&\{6,8\}&1\\
 5&44&0&0&0&\{5,6,7,8\}&1\\
 6&28&0&0&0&\{4,5,6,7\}&1\\
 7&18&0&0&0&\{4,5,6\}&5
\end{array}
\]
In addition, $a_{04}\geq1$. These are rational inequalities obtained directly by expansion of \eqref{bridge:classical-Bj}; every coefficient of index at least four not listed in $N_j$ is nonnegative. Using this table and the fact that  $0\leq x\leq1$, one can directly verify that
\begin{align*}
 B_0(x)&\geq \frac{97-257x+97x^2+108x^3}{2}\\
 &=54\left(x-\frac23\right)^2\left(x+\frac13\right)
    +\frac{205x^2-257x+81}{2}>0,\\
 B_1(x)&\geq74-88x+52x^2>0,\qquad
 B_2(x)\geq84-40x+47x^2>0,\\
 B_3(x)&\geq78,\quad B_4(x)\geq61,\quad
 B_5(x)\geq43,\quad B_6(x)\geq27,\quad B_7(x)\geq13.
\end{align*}
Thus, all $B_j$ are strictly positive on $[0,1]$. Since the weights in \eqref{bridge:classical-binomial} are nonnegative and sum to one, we conclude the proof of $Q>0$.
\end{proof}

\section{Self-adjoint realizations and graph cores}\label{app:domains}

In this appendix, we justify the domains of some auxiliary operators introduced in the main text.

\subsection{The axial auxiliary operator}\label{sec:H_kqj}

\begin{proof}[Proof of Proposition~\ref{ax:domain}]
It suffices to prove Proposition \ref{ax:domain} for $q=0$ because $H^{(3)}_{k,q,j}$ differs from $H^{(3)}_{k,0,j}$ by a bounded perturbation as seen in \eqref{ax:bounded-perturbation}. Importantly, fixing $q=0$ gives great convenience because, as seen in Steps 1-4 of Section \ref{sec:singular_transfer}, $H^{(3)}_{k,0,j}$ is linked to (a doublet of) $T_{E=0}$ by a unitary transformation.

{\color{blue}Step 1. Essential self-adjointness.}

To show that $H^{(3)}_{k,0,j}$ is essentially self-adjoint on $C_c^\infty(\Omega;\mathbb C^8)$, it suffices to prove that
\begin{equation} \label{eq:von_neumann_criteria_1}
\ker(H_{k,0,j,int}^{(3),*}\pm i)=\{0\}
\end{equation}
thanks to the von Neumann criterion (cf. \cite[Chapter V, Theorem 3.16 and Problem 3.17]{Kato1995}). Here $H_{k,0,j,int}^{(3),*}$ is the maximal operator introduced in Step 4 of Section \ref{sec:singular_transfer}, with $q=0$. Let $k=am$ and $\mathcal U_{j,0}$ be the unitary transformation in \eqref{ax:unitary-composition} for $E=0$. We use $C_0$ and $K$ as defined in \eqref{ax:doubled-resolvent}.
Since $T_0$ is compact, the block definition makes $C_0$ compact and self-adjoint on $\mathcal H_j\oplus\mathcal H_j$. Since $\mathcal U_{j,0}$ is unitary, $K$ is likewise compact and self-adjoint on $L^2(\Omega;\mathbb C^8)$. Suppose that $\Phi\in\operatorname{Dom}(H_{k,0,j,int}^{(3),*})$ satisfies
\begin{equation} \label{eq:von_neumann_criteria_2}
H_{k,0,j,int}^{(3),*}\Phi=i\Phi,
\quad
\Phi=\mathcal U_{j,0}(h,f)^{\top},
\quad (h,f)\in\mathcal H_j\oplus\mathcal H_j .
\end{equation}
Then we know that the spinors $\sqrt{r_B}h,\sqrt{r_A}f$ are tempered distributions, since $h,f\in L^2(\mathbb R^3;\mathbb C^4)$ and the multipliers have polynomial growth. They actually lift to $L^2$: first, the distributional identities obtained from \eqref{eq:von_neumann_criteria_2} and the coordinate calculation give
\begin{equation*}
D_m(\sqrt{r_B}h)=iw_Af,
 \quad D_m(\sqrt{r_A}f)=iw_Bh
 \quad\text{off the rotation axis}.
\end{equation*}
Each difference belongs to $H^{-1}_{\rm loc}(\mathbb R^3)$, since $\sqrt{r_B}h,\sqrt{r_A}f$ are locally square integrable and $w_X:L^2\to H^{-1}$ is bounded. The rotation axis is removable in this space. Indeed, given $0<\varepsilon<1$, let $\eta_\varepsilon(\rho)$ vanish for $\rho\leq\varepsilon^2$, equal one for $\rho\geq\varepsilon$, and be affine in $\log\rho$ in between. On each bounded cylinder, $\int|\nabla\eta_\varepsilon|^2=O(|\log\varepsilon|^{-1})$. Thus $\eta_\varepsilon\varphi\to\varphi$ in $H^1$ for every smooth compactly supported test spinor $\varphi$. By smooth approximation away from the axis, both differences vanish on $\eta_\varepsilon\varphi$ and hence on $\varphi$. The identities hold on $\mathbb R^3$. Since $D_m$ is invertible on tempered distributions and $D_m^{-1}:H^{-1}\to L^2$ is bounded, they imply
\begin{equation*}
\sqrt{r_B}h=iD_m^{-1}w_Af, \quad \sqrt{r_A}f=iD_m^{-1}w_Bh \quad\text{in }L^2(\mathbb R^3;\mathbb C^4).
\end{equation*}
Multiplication by $w_B$ and $w_A$, respectively, then gives
\begin{equation*}
h=iT_0f,\quad f=iT_0^*h.
\end{equation*}
These weighted compositions are justified by approximation of the sources in $L^2$, boundedness of $D_m^{-1}w_X$, $T_0$ and $T_0^*$, and closedness of multiplication by $w_X$. As a result,
\begin{equation*}
\Phi=iK\Phi \quad\text{in }L^2(\Omega;\mathbb C^8).
\end{equation*}
Then the self-adjointness of $K$ implies that $\Phi=0$ and hence completes the proof of \eqref{eq:von_neumann_criteria_1} for the eigenvalue $+i$; the $-i$ case is treated similarly.

{\color{blue}Step 2. Compact resolvent.}

Now we prove that the self-adjoint closure of $H_{k,0,j}^{(3)}$, temporarily denoted as $H$ for ease of notation, has compact resolvent. For an arbitrary $\Phi\in\operatorname{Dom}(H)$, write
\begin{equation*}
(h,f)^\top:=\mathcal U_{j,0}^*\Phi, \quad (h_0,f_0)^\top:=\mathcal U_{j,0}^*H\Phi.
\end{equation*}
Reversing the transformation and applying the same cutoff argument gives
\[
 D_m(\sqrt{r_B}h)=w_Af_0,\qquad D_m(\sqrt{r_A}f)=w_Bh_0
 \quad \text{in }\mathcal D'(\mathbb R^3;\mathbb C^4).
\]
Applying $D_m^{-1}$ and multiplying by the corresponding weights yields $h=T_0f_0$ and $f=T_0^*h_0$. Consequently,
\begin{equation} \label{eq:von_neumann_criteria_3}
\Phi=KH\Phi,
\quad\text{and hence}\quad
\|\Phi\|\leq\|K\|\,\|H\Phi\|
\qquad\text{for every }\Phi\in\operatorname{Dom}(H).
\end{equation}
This estimate shows that $H$ is injective and, since $H$ is closed, has closed range. Since we have shown that $H$ is self-adjoint, it holds that
\begin{equation*}
\operatorname{Ran}(H)^\perp=\ker(H^*)=\ker(H)=\{0\}.
\end{equation*}
Hence we conclude that $\operatorname{Ran}(H)=L^2(\Omega;\mathbb C^8)$ and the inverse $H^{-1}$ is bounded. Then \eqref{eq:von_neumann_criteria_3} implies $H^{-1}=K$, which concludes the proof by recalling that $K$ is compact.
\end{proof}

\subsection{The radial operator}\label{app:radial-domain}

\begin{proof}[Proof of Lemma~\ref{rad:domain}]

{\color{blue}Step 1. Essential self-adjointness.}

Note that $R$ is symmetric on $C_c^\infty((0,\infty);\mathbb C^8)$. Hence, to prove the essential self-adjointness of $R$, similar as in Step 1 of Appendix~\ref{sec:H_kqj}, it suffices to prove that the adjoint $R^{*}$ has no nonzero eigenvector with eigenvalue $z\in\{\pm i\}$ by the von Neumann criterion. Suppose that $Rf=zf$ in the sense of distribution with $f\in L^2$. Since the coefficients of $R$ are smooth away from $\mu=0$ (see \eqref{rad:operator}), we know $f^{\prime}$ is locally $L^2$, and hence $f$ is locally absolutely continuous. 

{\color{blue}Step 1.1.} We first investigate the behaviour of $f$ near zero. By definition of $R$, the equation $Rf=zf$ is rewritten as follows near $\mu=0$
\[
f^{\prime}=\frac{N}{2\mu}f+C_z(\mu)f,
\]
where $C_z$ is a bounded matrix-valued function. We first examine the component $f_-:=(I-N)f/2$. The preceding equation gives
\begin{equation} \label{eq:f_negative_near_zero_1}
(\mu^{1/2}f_-)^{\prime}= \mu^{1/2}\frac{I-N}{2}C_z(\mu)f.
\end{equation}
Since the right side is integrable near $\mu=0$, it follows that $\lim_{\mu\to 0^{+}}\mu^{1/2}f_-(\mu)$ exists, which necessarily equals zero (otherwise $f_-(\mu)\sim \mu^{-1/2}$ for small $\mu$, contradicting $f\in L^2$). Consequently, \eqref{eq:f_negative_near_zero_1} implies that
\begin{equation} \label{eq:f_negative_near_zero_2}
|f_-(\mu)| \leq C\mu^{-1/2}\int_0^\mu t^{1/2}|f(t)|\,dt
\leq C\sqrt\mu\,\|f\|_{L^2(0,\mu)} =o(\sqrt\mu).
\end{equation}
Next, let $f_+:=(I+N)f/2$. Since $f$ is square-integrable, so is $f_+$, by which we know there exists a sequence $\mu_n\to 0^{+}$ such that
\begin{equation} \label{eq:f_negative_near_zero_3}
\mu_n|f_+(\mu_n)|^2\to 0,
\end{equation}
otherwise the comparison test leads to a contradiction to the integrability.

{\color{blue}Step 1.2.} We now use \eqref{eq:f_negative_near_zero_2}-\eqref{eq:f_negative_near_zero_3} to show that $f=0$. This is achieved by considering the current associated with the radial operator $R$
\[
\mathcal I(\mu):=f(\mu)^*Af(\mu),\quad \text{with }A:=\zeta_3\sigma_1.
\]
Note that the matrices $A$ and $N$ are Hermitian involutions and satisfy $\{A,N\}=0$. In particular, using the equation $Rf=zf$ and the Hermitian potential in $R$, one can check that
\begin{equation}\label{rad:current}
\mathcal I^{\prime}(\mu)=-2\operatorname{Im}z\,|f(\mu)|^2.
\end{equation}
As a consequence, since $f\in L^2$, the current $\mathcal I(\mu)$ has finite limits at both endpoints $\mu=0$ and $\mu=\infty$. Moreover, $A$ interchanges the two eigenspaces of $N$, by which one can verify that
\[
|\mathcal I(\mu)|\leq2|f_+(\mu)|\,|f_-(\mu)|.
\]
Hence, using \eqref{eq:f_negative_near_zero_2}-\eqref{eq:f_negative_near_zero_3}, we know $\mathcal I(\mu_n)\to0$ and hence $\mathcal I(0)=0$. Since $f\in L^2$, there is a sequence tending to infinity along which $|f|\to0$. The limit $\mathcal I(\infty)$ is therefore zero, and \eqref{rad:current} implies
\[
0=\mathcal I(\infty)-\mathcal I(0)
=-2\operatorname{Im}z\,\|f\|^2.
\]
Since $\operatorname{Im}z=\pm 1\neq 0$, it follows that $f=0$. This proves that $R$ is essentially self-adjoint.

Finally, since the adjoint $R^{*}$ has the maximal domain (the right side of \eqref{rad:maximal-domain}) and the essential self-adjointness identifies $R^{*}$ with the self-adjoint closure, we conclude the proof of the claimed domain \eqref{rad:maximal-domain}.

{\color{blue}Step 2. Compactness of the graph-domain embedding.}

To show the compact embedding under the graph norm, we require a coercive estimate of $\|Rf\|^2+C\|f\|^2$. We use the expression for $R^2$ in \eqref{rad:square}.
Near $\mu=0$, it has the leading-order asymptotics 
\begin{equation*}
R^2= -\partial_\mu^2+ (\tfrac14I-\tfrac12N)\mu^{-2} + r_0(\mu)
\end{equation*}
where the coefficient matrix $\tfrac14I-\tfrac12N$ has eigenvalues $-1/4$ and $3/4$, and the remainder $r_0(\mu)$ is bounded near zero. On the other hand, as $\mu\to\infty$, the term $k^2\sinh^2\mu$ dominates the others. Consequently, for some $C=C(k,q)>0$,
\begin{equation}\label{rad:compact-form}
\|Rf\|^2+C\|f\|^2
\geq \int_0^\infty
\left( |f'|^2-\frac{|f|^2}{4\mu^2} +\frac{k^2}{2}\sinh^2\mu\,|f|^2 \right)d\mu
\end{equation}
for every $f\in C_c^\infty((0,\infty);\mathbb C^8)$. Note that the critical Hardy form on the right side is nonnegative: writing $f=\sqrt\mu\,g$, integration by parts gives
\[
\int_0^\infty
\left(|f'|^2-\frac{|f|^2}{4\mu^2}\right)d\mu
=
\int_0^\infty\mu|g'|^2\,d\mu,
\quad
\|f\|^2=\int_0^\infty\mu|g|^2\,d\mu.
\]
After multiplication by a fixed cutoff supported on $[0,M]$, these integrals are the radial $H^1$ and $L^2$ quantities on a two-dimensional disk, up to a constant factor. The compact embedding of $H^1$ into $L^2$ on that disk therefore gives local compactness, including at $\mu=0$. Outside $(0,M)$, the last term in \eqref{rad:compact-form} controls the
tail:
\[
\int_M^\infty|f|^2\,d\mu
\leq
\frac{2}{k^2\sinh^2M}
\bigl(\|Rf\|^2+C\|f\|^2\bigr).
\]
Given a sequence $(f_n)$ bounded in the graph norm of $R$, choose $\varphi_n\in C_c^\infty((0,\infty);\mathbb C^8)$ with
\[
\|f_n-\varphi_n\|+\|R(f_n-\varphi_n)\|\leq n^{-1}.
\]
The preceding estimates give an $L^2$-convergent subsequence of $\varphi_n$, and hence of $f_n$. This proves the compactness of the graph-domain embedding.
\end{proof}


\subsection{\texorpdfstring{The separated operator in \eqref{rad:decomposition}}{The separated operator}}\label{app:separated-domain}

\begin{proof}[Proof of Lemma~\ref{rad:sum-domain}]

\text{}

{\color{blue}Step 1.} We first prove the essential self-adjointness of the angular operator $A_0$. To this end, we apply the current argument in Step 1 of Appendix~\ref{app:radial-domain} for $A_0$ with the two end points being $\nu=0$ and $\nu=\pi$. Here we only point out the necessary changes in the argument and skip the details. Suppose $A_0f=zf$. Then near $\nu=0$ it holds that 
\[
f^{\prime}=\frac{\widehat N}{2\nu}f+C_z(\nu)f,
\]
with $C_z$ being bounded and $\widehat N:=-iST$ being a Hermitian involution that anticommutes with $S$ (recall the definition of $S,T$ in \eqref{rad:angular-operator}). Near $\nu=\pi$, by the change of variables $\nu\to \pi-\nu$, a similar equation holds with the corresponding change $\widehat N\to -\widehat N$. Thus, the indicial exponents are $\pm1/2$ in \eqref{eq:f_negative_near_zero_1} at both endpoints, with the roles of the two eigenspaces of $\widehat N$ interchanged at $\nu=\pi$. On the other hand, here we consider the current
\begin{equation*}
\mathcal I(\nu):=f(\nu)^*Sf(\nu) .
\end{equation*}
Then, following the same argument as in Lemma~\ref{rad:domain}, one can show that $\mathcal I(\nu)$ vanishes at both endpoints. Since its derivative equals $-2\operatorname{Im}z\,|f|^2$, we then conclude that $f\equiv 0$ for $z=\pm i$. Consequently, $A_0$ is essentially self-adjoint, and its closure has the maximal distributional domain.

{\color{blue}Step 2.} We next prove that the spectral projections of $R^2$ in Proposition \ref{rad:main} commute with $A_0$. First, by \eqref{rad:matrix-relations}, we know the matrices $S,T$ preserve $\Dom(R)$ and anti-commute with the operator $R$; in particular, these matrices commute with every spectral projection of $R^2$. To extend this commutation relation to the full operator $A_0$, we note that the radial spectral projection acts only on the variable $\mu$ and hence commutes both with the differential $\partial_{\nu}$ and the multiplier $\csc \nu$. Consequently, recalling the definition of $A_0$ in \eqref{rad:angular-operator}, we know any such spectral projection preserves the maximal domain of $A_0$ and commutes with $A_0$. In particular, these projections also commute with the resolvent of $A_0$.

{\color{blue}Step 3.} We now prove \eqref{rad:graph-identity}. Since $R$ has compact resolvent, the spectrum of $R^2$ consists of discrete eigenvalues with finite multiplicity with the associated eigenfunctions forming an orthogonal basis of the radial Hilbert space. In particular, each eigenspace is invariant under $R,S,T$ and reduces $A_0$. On the block $L^2((0,\pi);\mathcal E_\lambda)$, where $\mathcal E_\lambda:=\ker(R^2-\lambda)$, the radial operator becomes a bounded Hermitian matrix $R_\lambda$, with $\|R_\lambda\|\leq\sqrt{\lambda}$. The restriction $A_{0,\lambda}$ of $A_0$ is self-adjoint because this block reduces $A_0$. Consequently, $H_\lambda:=R_\lambda+A_{0,\lambda}$ is self-adjoint on $\operatorname{Dom}(A_{0,\lambda})$ by the bounded perturbation theorem. Moreover, on each block, the anti-commutation relations $\{S,R\}=\{T,R\}=0$ imply that the mixed terms vanish:
\[
\|(R+A_0)f\|^2=\|Rf\|^2+\|A_0f\|^2 .
\]
Taking the orthogonal sum of these self-adjoint block operators defines a self-adjoint operator $H_0$ on $L^2(\Omega;\mathbb{C}^8)$. The above identity also shows that $\Dom(H_0)=\Dom(R)\cap\Dom(A_0)$ and proves \eqref{rad:graph-identity}. It also identifies the square $H_0^2$ with the joint spectral sum $R^2+A_0^2$.

{\color{blue}Step 4.} Finally, we show that $C_c^\infty(\Omega;\mathbb C^8)$ is a graph core of $H_0$. This is achieved by consecutive approximations by compactly supported functions in radial and angular variables. Let $f\in\operatorname{Dom}(H_0)$, and write
\[
\|g\|_{H_0}^2
:=\|g\|^2+\|H_0g\|^2
=\|g\|^2+\|Rg\|^2+\|A_0g\|^2.
\]
We first truncate the radial spectral decomposition. Put $E_n:=\mathbf 1_{[0,n]}(R^2)$ and $f_n:=E_nf$. Since $E_n$ commutes with $H_0$, we know
\[
\|f_n-f\|_{H_0}^2
=\|(E_n-I)f\|^2+\|(E_n-I)H_0f\|^2
\longrightarrow0.
\]
For fixed $n$, let $\{e_r\}_{r=1}^{d_n}$ be an orthonormal basis of the radial space $\operatorname{Ran}E_n$. On this finite-dimensional radial space the restrictions of $S$ and $T$ are anticommuting Hermitian involutions. The endpoint argument in Step~1 therefore gives $C_c^\infty((0,\pi);\operatorname{Ran}E_n)$ as a graph core for the restricted angular operator. Thus there are functions
\[
g_\ell(\mu,\nu)
=\sum_{r=1}^{d_n}e_r(\mu)a_{\ell r}(\nu),
\quad
a_{\ell r}\in C_c^\infty((0,\pi)),
\]
such that
\[
\|g_\ell-f_n\|+\|A_0(g_\ell-f_n)\|\longrightarrow 0.
\]
Moreover,
\[
\|g_\ell-f_n\|_{H_0}^2
\leq(1+n)\|g_\ell-f_n\|^2
+\|A_0(g_\ell-f_n)\|^2
\longrightarrow0.
\]
Next, choose $e_{r,m}\in C_c^\infty((0,\infty);\mathbb C^8)$
such that, for $\delta e_{r,m}:=e_{r,m}-e_r$,
\[
\|\delta e_{r,m}\|+\|R\delta e_{r,m}\|
\longrightarrow0 .
\]
Then we define
\[
g_{\ell,m}(\mu,\nu)
:=\sum_{r=1}^{d_n}e_{r,m}(\mu)a_{\ell r}(\nu)
\in C_c^\infty(\Omega;\mathbb C^8).
\]
This gives the desired approximation: for fixed $\ell$, we have
\[
\begin{aligned}
\|g_{\ell,m}-g_\ell\|_{H_0}
\leq{}&\sum_{r=1}^{d_n}
\bigl(\|a_{\ell r}\|+\|a_{\ell r}'\|
+\tfrac12\|a_{\ell r}\csc\nu\|\bigr)
\|\delta e_{r,m}\|\\
&+\sum_{r=1}^{d_n}\|a_{\ell r}\|\,\|R\delta e_{r,m}\|
\longrightarrow0.
\end{aligned}
\]
A diagonal choice therefore proves that
$C_c^\infty(\Omega;\mathbb C^8)$ is a graph core for $H_0$.

Finally, we remark that since $kW$ is bounded and self-adjoint, $H_0+kW$ is self-adjoint with the domain
\[
\operatorname{Dom}(H_0+kW)=\operatorname{Dom}(H_0)
\]
and the graph norm $\|f\|_{H_0+kW}$ is equivalent to $\|f\|_{H_0}$. Thus the same test functions form a graph core for $H_0+kW$. By \eqref{rad:decomposition}, its restriction to this core agrees with $H^{(3)}_{k,q,1/2}$. The essential self-adjointness of $H^{(3)}_{k,q,1/2}$ in Proposition \ref{ax:domain} therefore gives $H_0+kW=H^{(3)}_{k,q,1/2}$.
\end{proof}

\bibliographystyle{amsplain}
\bibliography{ref-final}
\end{document}